\documentclass[useAMS,usenatbib]{mn2e}

\usepackage{graphicx}
\usepackage{amsmath}
\usepackage{amssymb}
\usepackage{float}
\usepackage{multirow}
\usepackage{url}
\usepackage{longtable,lscape}

\newcommand{\nH}{n_{\rm _{H}} }

\newcommand{\nHI}{n_{\rm _{HI}} }

\newcommand{\NHI}{N_{\rm HI}}

\newcommand{\Msun}{{\rm M_{\odot}} }

\newcommand{\Mpch}{h^{-1} {\rm Mpc} }

\newcommand{\owls}{{\sc owls} }

\newcommand{\gadget}{ \textsc{gadget} }
\newcommand{\arepo}{ \textsc{arepo} }

\newcommand{\ion}[2]{#1{\small\rm{#2}}\relax}

\newcommand{\HI}{ \ion{H}{I} }

\newcommand{\fN}{\ensuremath{f(\NHI)}}

\newcommand{\plotphrase}{{\em Top panel}: $\fN$ for models with ({\em solid lines}) and without ({\em dashed lines}) a correction for H$_2$.  {\em Grey points with error bars} are the observational data from Noterdaeme et al. (2012) and the {\em solid grey line} represents the double power-law fit of Noterdaeme et al. (2009).  {\em Lower pane}l: $\log_{10} (f_{\rm MODEL} / f_{\rm REF}$).  The {\em grey regions} are the same as in Fig. \ref{fig:main_cddf_big}}

\title[Impact of physical processes on simulated LLSs and DLAs]{The impact of  different physical processes on the statistics of Lyman-limit and damped Lyman-$\alpha$ absorbers}

\author[Gabriel Altay et al.]
       {Gabriel Altay$^{1,2}$\thanks{E-mail: gabriel.altay@gmail.com}, 
         Tom Theuns$^{1,3}$,
         Joop Schaye$^{4}$,
         C. M. Booth$^{5}$,
         Claudio Dalla Vecchia$^{6}$ \\
         $^{1}$Institute for Computational Cosmology, Department of
         Physics, University of Durham, South Road, Durham DH1 3LE \\
         $^{2}$Center for Relativistic Astrophysics, School of Physics, 
         Georgia Institute of Technology, 837 State Street, Atlanta, GA, USA \\         
         $^{3}$Department of Physics, University of Antwerp, Campus
         Groenenborger, Groenenborgerlaan 171, B-2020 Antwerp, Belgium \\
         $^4$Leiden Observatory, Leiden University, P.O. Box 9513, 2300
         RA Leiden, the Netherlands \\
         $^5$Department of Astronomy \& Astrophysics, 
         The University of Chicago, Chicago, IL, 60637 \\
         $^{6}$Max Planck Institute for Extraterrestrial Physics, 
         Gissenbachstra\ss{}e, 85748 Garching, Germany 
       }

\begin{document}

\date{Accepted 201? ???? ??.
      Received 201? ???? ??;
      in original form 2010  xx}

\pagerange{\pageref{firstpage}--\pageref{lastpage}}
\pubyear{201?}
\maketitle

\label{firstpage}

\begin{abstract}
We compute the $z=3$ neutral hydrogen column density distribution function $\fN$ for 19 simulations drawn from the \owls project using a post-processing correction for self-shielding calculated with full radiative transfer of the ionising background radiation.  We investigate how different physical processes and parameters affect the abundance of Lyman-limit systems (LLSs) and damped Lyman-$\alpha$ absorbers (DLAs) including: {\em i}) metal-line cooling; {\em ii}) the efficiency of feedback from SNe and AGN; {\em iii}) the effective equation of state for the ISM; {\em iv}) cosmological parameters; {\em v}) the assumed star formation law and; {\em vi}) the timing of hydrogen reionization .  We find that the normalisation and slope, $\mathcal{D} = d \, \log_{10} f / d\, \log_{10} \NHI$, of $\fN$ in the LLS regime are robust to changes in these physical processes.  Among physically plausible models,  $\fN$ varies by less than 0.2 dex and $\mathcal{D}$ varies by less than 0.18 for LLSs.  This is primarily due to the fact that these uncertain physical processes mostly affect star-forming gas which contributes less than 10\% to $\fN$ in the LLS column density range. At higher column densities, variations in $\fN$ become larger (approximately 0.5 dex at $\NHI = 10^{22} {\rm cm^{-2}}$ and 1.0 dex at $\NHI = 10^{23} {\rm cm^{-2}}$) and molecular hydrogen formation also becomes important.  Many of these changes can be explained in the context of self-regulated star formation in which the amount of star forming gas in a galaxy will adjust such that outflows driven by feedback balance inflows due to accretion.  Tools to reproduce all figures in this work can be found at the following url: \url{https://bitbucket.org/galtay/hi-cddf-owls-1}
\end{abstract}


\section{Introduction}
Hydrogen is the most abundant element in the Universe and an excellent tracer of cosmic structure.  Neutral hydrogen can be detected as Lyman-$\alpha$ (and higher-order Lyman series) absorption lines in the spectra of bright ultraviolet (UV) sources.  The number density and physical cross-section of absorbers together determine the \ion{H}{I} column density distribution function\footnote{The number of lines per unit absorption distance, per unit column density, $\fN$, as defined in Eq.~\ref{eq:fhi} below.}(CDDF), $\fN$.
Over the past several decades, ground-based spectroscopic surveys have led to an increasingly accurate observational determination of $\fN$  
\cite[][]{Carswell_84, Tytler_87, Lanzetta_91, Petitjean_93, Storrie-Lombardi_00, Peroux_01, Kim_02, Prochaska_05, Prochaska_10, Omeara_07, Omeara_13, Noterdaeme_09, Ribaudo_11_Census, Noterdaeme_12, Rudie_13, Kim_13, Zafar_13, Patra_13}.  

The nature of the Lyman-$\alpha$ transition allows for an estimate of the abundance of absorbers with column densities between $\NHI \approx 10^{12} {\rm cm^{-2}}$ and $\NHI \approx 10^{22} {\rm cm^{-2}}$.  Historically, this column density range has been divided into three groups: absorbers with column densities below $\NHI = 10^{17.2} {\rm cm^{-2}}$ or the Lyman-$\alpha$ forest; those with column densities above $\NHI = 10^{20.3} {\rm cm^{-2}}$ or Damped Lyman-$\alpha$ systems (DLAs); and those with intermediate column densities or Lyman Limit Systems (LLSs). Relevant reviews can be found in \cite{Rauch_98}, \cite{Wolfe_05} and \cite{Meiksin_09}.

The opacity of the atmosphere at UV wavelengths makes it impossible to use ground-based surveys to detect absorbers at redshifts less than $z \approx 1.7$, but the new {\em Cosmic Origins Spectrograph}\footnote{\url{www.stsci.edu/hst/cos}} on the {\em Hubble Space Telescope} provides significant capacity to probe low redshift systems \citep[e.g.,][]{Battisti_12}. While low $z$ absorption line observations will compliment studies of neutral hydrogen in emission \cite[e.g.,][]{Duffy_12}, the bulk of observed absorption lines are currently at high $z$.  

Observations of LLSs and DLAs probe gas in and around galaxies and therefore can be used to test model predictions of galaxy formation theories.  In current models, feedback from star formation plays a crucial role in the suppression of low-mass galaxy formation, the maintenance of low star formation efficiencies, and the formation of spiral galaxies with low bulge-to-disk ratios \citep[see][for a recent review]{Benson_10}.  Galactic outflows generated by feedback have been detected at both low \citep[e.g.,][]{Heckman_90} and high \citep[e.g.,][]{Pettini_01} redshift, and high-resolution simulations of the interstellar medium (ISM) of galactic disks suggest that supernovae (SNe) can indeed power strong outflows \citep[e.g.,][]{Creasey_13}.  

The detection of metal line absorption coincident with Lyman-$\alpha$ forest lines \citep[e.g.,][]{Cowie_95, Schaye_03, Aguirre_08} indicates that these outflows transport material into the intergalactic medium (IGM).  
However, numerical simulations suggest that the impact of outflows on Lyman-$\alpha$ forest flux statistics is small \citep[e.g.,][]{Viel_13}.  This is likely due to the tendency of galactic outflows to travel the path of least resistance through under-dense regions as opposed to the denser gas responsible for \ion{H}{I} absorbers \citep[e.g.,][]{Theuns_02, Brook_11}.

Given that higher column density systems are more closely associated with galaxies \citep[e.g.,][]{Steidel_10, Ribaudo_11_ColdAcc, VanDeVoort_12, Rakic_12, Rudie_13}, it is reasonable to ask if the signature of outflows is visible in the abundance of LLSs and DLAs.  In this paper we examine 19 different models from the \owls\  suite of cosmological hydrodynamical simulations \citep{Schaye_10} which employ a variety of sub-grid implementations, some of which directly relate to the way galactic outflows are driven.  The goal is to isolate which physical processes are most important in determining the properties of absorbers and to reduce the large amount of freedom currently available in sub-grid implementations of feedback.   

One difficulty not encountered when working with the Lyman-$\alpha$ forest is that LLSs and DLAs are dense enough to self-shield from the hydrogen ionising background, a phenomenon not included in most hydrodynamical simulations. Pioneering numerical work on the abundance of dense \ion{H}{I} absorbers was presented in \cite{Katz_96} and \cite{Haehnelt_98}.  In these works, self-shielding was modelled with 1-D radiative transfer calculations or correlations between neutral column density and total volume density.   In the past several years, computing hardware and algorithms have advanced to the point where it is feasible to incorporate full 3-D radiative transfer to calculate self-shielding \citep{Kohler_07, Pontzen_08, Altay_11, McQuinn_11, Fumagalli_11, Erkal_12, Cen_12, Yajima_12, Rahmati_13_str, Rahmati_13_evo}.  In this work, we make use of a 3-D radiative transfer code called {\sc urchin} \footnote{\url{https://bitbucket.org/galtay/urchin}}\citep{Altay_13}, which is specifically designed to model self-shielding in the post-reionisation universe.

This paper is organised as follows. In \S2 we briefly summarise the different sub-grid models included in the \owls\ suite, and our method for using the code {\sc urchin} to calculate the abundance of neutral hydrogen. In \S3 we introduce the HI CDDF and discuss its shape in the LLS regime.  In \S4 we describe the incidence of absorption systems. In \S5 we discuss the DLA column density range in a general way, describe what role self-regulation plays in shaping $\fN$ and present the cosmic density of HI from DLAs in the different \owls models.  In \S6 we examine the DLA range of each \owls model in detail and in \S7 we present our conclusions.

\begin{table*}
 \centering
 \caption{Galactic wind and cooling models discussed in this work.  The symbol $\xi$ represents energy injected per unit stellar mass formed and is proportional to $\eta v_{\rm w}^2$.   In model {\sc ref},  $v_{\rm w} = v_{\rm w}^* = 600$ km s$^{-1}$ and $\eta=\eta^*=2$.  In all models discussed here, $\nH^* = 10^{-1} {\rm cm^{-3}}$.  To simplify notation we make use of the arbitrary constant $\sigma_0 = 150$ km s$^{-1}$.  }
  \begin{tabular}{lllll}

    \hline
    \hline
    Model & 
    $v_{\rm w}$ & 
    $\eta$ & 
    $\xi / \xi_{\rm REF}$ &
    Notes 
    \\
     & 
    [km s$^{-1}$] & 
    [$\dot{M}_{\rm wind} / \dot{M}_{*}$] & 
     &  
    \\
    \hline
    \hline

    {\sc wml1v848} & 848 & 1 & 1.0  &  \\
    {\sc ref}      & 600 & 2 & 1.0  &  \\
    {\sc wml4v424} & 424 & 4 & 1.0  &   \\
    {\sc wml8v300} & 300 & 8 & 1.0  &  \\
    {\sc wml4}     & 600 & 4 & 2.0  &  \\
    
    {\sc dblimfcontsfv1618} & 600, 1618 & 2 & 1.0, 7.3  & 
    High velocity for top heavy IMF  \\

    {\sc wvcirc} & 5 $\sigma$ & $\sigma_0 / \sigma$ & 
    $(25/32) \, \sigma / \sigma_0$ &
    Momentum driven winds, $\sigma = v_{\rm c} \sqrt{2}$  \\

    {\sc wdens} & $600 (\nH/\nH^*)^{1/6}$ & 
    $2 (v_{\rm w}/v_{\rm w}^*)^{-2}$ & 1.0  &  
    $v_{\rm w} \propto c_{\rm s,eos}$ \\

    {\sc nozcool} & 600 & 2 & 1.0  & No metal line cooling  \\
    {\sc nosn\_nozcool} & - & - & - & 
    No SNe feedback, no metal line cooling  \\
    \hline

\label{tab:models_feedbk}
\end{tabular}
\end{table*}

\section{Optically Thick Absorbers in {\sc owls}}

The OverWhelmingly Large Simulations \citep[{\sc owls},][]{Schaye_10} consist of a suite of cosmological hydrodynamical simulations performed using a version of the Smoothed Particle Hydrodynamics (SPH) code \gadget last described in \cite{Springel_05}.  The version used has been modified to include sub-grid routines that model unresolved physics, such as star formation, feedback from both stars and accreting black holes, radiative cooling and chemodynamics.  The suite consists of a reference model ({\sc ref}) and a group of models in which the sub-grid physics implementations are systematically varied.  We begin with a brief overview of the \owls models.  Pertinent model details can be found in Tables \ref{tab:models_feedbk} and \ref{tab:models_other}.  For a more complete description, we refer the reader to \cite{Schaye_10}.

\subsection{The \owls sub-grid models}
The \owls\ reference model, {\sc ref}, assumes a $\Lambda$CDM cosmological model with parameters taken from the Wilkinson Microwave Anisotropy Probe 3-year results \citep[WMAP3,][]{2007ApJS..170..377S}, \{$\Omega_{\rm m}, \Omega_{\rm b}, \Omega_{\Lambda}, \sigma_8, n_s, h$\} = \{0.238, 0.0418, 0.762, 0.74, 0.951, 0.73\}, and a primordial helium mass fraction of $Y=0.248$.  Because we are mostly interested in comparisons between \owls models, the actual values of the assumed cosmological parameters are not crucial. However, we note that the WMAP3 values are all within $\approx$ 10\% of the WMAP7 values \citep{Komatsu_11}.  The most significant difference is in $\sigma_8$ and we will demonstrate that its value does impact the statistics of optically thick absorbers.   All simulations examined in this work evolved 512$^3$ SPH particles and an equal number of dark matter particles in a cubic volume 25 comoving $h^{-1}$~Mpc on a side.  These choices lead to an initial baryonic particle mass of $1.4\times 10^6 h^{-1} \Msun$.  An equivalent Plummer softening for gravitational forces of 2 co-moving $h^{-1}$~kpc was used at high redshifts.  Once the gravitational softening reached 0.5 physical $h^{-1}$~kpc it was fixed at that value.  All simulations considered in this work were evolved from redshift $127$ down to redshift $2$, but we only consider the $z=3$ outputs. 

\subsubsection{Star formation} 
Gas above a density of $\nH = 10^{-2} - 10^{-1} {\rm cm^{-3}} $ is susceptible to thermo-gravitational instabilities and is expected to be multi-phase and star forming \citep{Schaye_04}.  Star formation is implemented in \owls\ by moving gas particles with $\nH \ge \nH^* = 10^{-1} {\rm cm^{-3}}$ onto a polytropic equation of state with $p\propto\rho^{\gamma_{\rm eos}}$ which represents the multi-phase ISM.   These gas particles are converted probabilistically into collisionless star particles at a rate determined by gas pressure \citep{Schaye_08}. Such an implementation guarantees that simulated galaxies follow a Kennicutt-Schmidt law \citep{Kennicut_98},
\begin{equation}
\dot{\Sigma}_{\rm SFR} = A_{\rm ks} \, 
(\Sigma_{\rm g}/ \Msun\,{\rm pc}^{-2})^{n_{\rm ks}}\,.
\label{eq:KS}
\end{equation}
In the reference model we use  $A_{\rm ks} = 1.5 \times 10^{-4} \, h^{-1} \Msun \, {\rm yr}^{-1} \, {\rm kpc}^{-2}$, $n_{\rm ks}=1.4$, and $\gamma_{\rm eos}=4/3$.

\subsubsection{Stellar feedback} 
The efficiency of stellar feedback depends on the energy injected into the surrounding gas per unit stellar mass formed, $\xi$.  This energy can be used to either heat neighbouring gas particles or launch them into a wind.  The remaining energy fraction, $1-\xi$, is assumed to be lost due to unresolved radiative cooling.  

The kinetic feedback model used in \owls is fully described in  \cite{DallaVecchia_08} and is parameterised by the wind launch velocity $v_{\rm w}$ and the initial wind mass loading, $\eta = \dot M_{\rm w} / \dot M_\star$ where $\dot M_\star$ is the star formation rate and $\dot M_{\rm w}$ is the rate at which mass is added to the wind.  In this case, $\xi$ is related to the feedback parameters as,
\begin{equation}
{1\over 2}\,\eta\,v_{\rm w}^2 = \xi \, \epsilon_{\rm SN}\,.
\end{equation}
where $\epsilon_{\rm SN}$ is the amount of SNe energy released per unit stellar mass formed. Assuming the IMF of \cite{Chabrier_03} for stars with masses in the range $0.1-100 \, \Msun$, and that stars in the mass range  $6-100 \, \Msun$ end their lives as core-collapse SNe each releasing $10^{51}$~erg, the appropriate value for $\epsilon_{\rm SN}$ is approximately $1.8\times 10^{49}~{\rm erg}\,\Msun^{-1}$. The default model  ({\sc ref}) has $\eta=2$ and $v_{\rm w}=600$~km~s$^{-1}$ and hence $\xi=0.4$, implying 40\% of the core collapse SN energy is used to drive a wind for a \cite{Chabrier_03} IMF. In general, 
\begin{eqnarray}
\xi &=& 0.4\,{\eta\over 2}\,\left({v_{\rm w}\over 600~{\rm km}~{\rm s}^{-1}}\right)^2 
\end{eqnarray}
We report the value of $\xi$ for all {\sc owls} models considered in this work in Table~\ref{tab:models_feedbk}.

\subsubsection{Stellar evolution} 
Assuming a star particle represents a single stellar population with chemical abundances taken from its parent gas particle, we follow the timed release of 11 elements by AGB stars, and both type~Ia and type~II supernovae \citep{Wiersma_09_chem}. The {\sc ref} model assumes the stellar initial mass function of \cite{Chabrier_03}. Elements released by evolving stars are spread to neighbouring gas particles weighted by the SPH smoothing kernel.

\subsubsection{Radiative cooling and heating} 
Radiative cooling rates along with photo-heating due to an imposed evolving UV/X-ray background are calculated element-by-element
using the publicly available photo-ionization package {\small CLOUDY} \citep[last described in][]{Ferland_89} as described in  \cite{Wiersma_09_cool}.  All \owls simulations which include a UV background  use the \cite{HM01} model.  Model {\sc ref} assumes \ion{H}{I} reionisation occurs instantaneously at redshift $z_{\rm reion}=9$.  The reionisation of \ion{He}{II} is modelled by injecting two eV per atom at redshift $z\approx 3.5$ in order to match the \cite{Schaye_00} IGM temperature measurements \citep[see][]{Wiersma_09_cool}.  

\subsubsection{Variations on the reference model} 
A full description of all \owls model variations can be found in \cite{Schaye_10}. Visual representations of the physical conditions in and around high-redshift galaxies in several \owls models can be found in \cite{VanDeVoort_Schaye_12}.  Here we provide a brief description of those models which involve cooling or feedback.  Model {\sc nozcool} did not include cooling from metals while model {\sc nosn\_nozcool} includes neither energetic stellar feedback or cooling from metals.  Model {\sc wml4} used twice the {\sc ref} feedback energy per unit stellar mass formed, $\xi$, by doubling the mass loading ($\eta=4$, $v_w=600$~km~s$^{-1}$) while model {\sc mill} used the \lq Millennium cosmology\rq\, \{$\Omega_{\rm m}, \Omega_{\rm b}, \Omega_{\Lambda}, \sigma_8, n_s, h$\} = \{0.25, 0.045, 0.75, 0.9, 1, 0.73\} and the same feedback parameters as {\sc wml4}.  Model {\sc dblimfcontsfv1618} ({\sc dblimf} hereafter) used more effective feedback ($\eta=2$, $v_{\rm w}=1618$ km s$^{-1}$) in high pressure gas approximating a top heavy IMF in star-bursting galaxies. Models {\sc wml1v848}, {\sc wml4v424}, and {\sc wml8v300} vary wind mass loading and launch velocity while keeping $\xi$ the same as in {\sc ref}.  

In models {\sc wdens} and {\sc wvcirc} the stellar feedback parameters $\eta$ and $v_{\rm w}$ scale with galaxy properties.  In {\sc wdens} the launch velocty scales with the local sound speed of the star forming gas, $v_{\rm w} \propto c_{\rm s, eos} \propto \nH^{1/6}$, and the mass loading is such that a constant amount of energy is injected per unit stellar mass formed, $\eta \propto v_{\rm w }^{-2}$ (i.e., $\xi=\xi_{\rm REF}$).  In {\sc wvcirc} the launch velocity scales with the circular velocity of the host galaxy, $v_{\rm w} \propto v_{\rm c}$, and the mass loading is such that a constant amount of momentum is injected per unit stellar mass formed, $\eta \propto v_{\rm w}^{-1}$, approximating momentum-driven winds. Finally, model {\sc agn} includes feedback from accreting black holes using the implementation described in \cite{Booth_09}.  For convenience, we provide tables of the relevant parameters for variations involving stellar feedback and cooling (Table~\ref{tab:models_feedbk}) and variations that do not (Table~\ref{tab:models_other}).

\subsubsection{Resolution and Box Size}

In addition to the physics variations described above, the reference model was run with a variety of  box sizes and numerical resolutions.  Simulations with small box sizes will not contain halos above a given mass and therefore may be deficient in \ion{H}{I} absorbers.  The number of particles used to represent the density field also plays an important role in determining the halo population.  The mass below which halos are unresolved and the internal structure of higher mass halos below some length scale is determined by the resolution.  In addition to the consequences for the halo population, the resolution can also influence the density field used for the radiative transfer algorithm described below.  

We have analyzed variations around our chosen box size and resolution, 25 $\Mpch$ with $2 \times 512^3$ particles, using box sizes ranging from 6 $\Mpch$ to 100 $\Mpch$ and mass (spatial) resolutions differing by factors of 64 (4).   While these variations do play a role, changes in the normalization of the column density distribution function are limited to approximately 0.25 dex.
\cite{Rahmati_13_evo} also examined these issues using an independent radiative transfer code (see their Fig. B1) and found similar results.   As we are interested in relative changes between \owls model variations that involve physical effects, we focus our attention on a series of models with fixed resolution and box size.

\begin{table*}
  \centering
  \caption{Variations on model {\sc ref} other than those involving galactic winds and cooling used in this work.  All models considered were run in a 25 $\Mpch$ box with $512^3$ dark matter and $512^3$ baryonic particles.  The variations cover the ISM effective equation of state exponent $\gamma_{\rm eos}$, the timing of reionization, $z_{\rm reion}$, the amplitude, $A_{\rm ks}$, and slope, $n_{\rm ks}$, of the Kennicut-Schmidt law, feedback from AGN, and cosmological parameters.     }
  \begin{tabular}{ll}
    \hline
    \hline
    Model & Notes  \\
    \hline
    \hline
    {\sc ref} & WMAP 3, $z_{\rm reion} = 9$, $\gamma_{\rm eos}$ = 4/3, $A_{\rm ks} = 1.5 \times 10^{-4} \, h^{-1} M_\odot \, {\rm yr}^{-1} \, {\rm kpc}^{-2}$, $n_{\rm ks}=1.4$ \\
    {\sc eos1p0} & Slope of effective EOS changed to $\gamma_{\rm eos}$ = 1 (isothermal) \\
    {\sc eos1p67} & Slope of effective EOS changed to $\gamma_{\rm eos}$ = 5/3 (adiabatic) \\
    {\sc noreion} & HM01 UV background not present\\
    {\sc reionz06} & HM01 UV background turned on at $z_{\rm reion} = 6$ \\
    {\sc reionz12} & HM01 UV background turned on at $z_{\rm reion} = 12$ \\
    {\sc sfslope1p75} & $n_{\rm ks}$ increased from 1.4 to 1.75 \\
    {\sc sfamplx3} & $A_{\rm ks}$ increased by a factor of 3 \\
    {\sc agn} & Includes AGN as well as SNe feedback \\ 
    {\sc mill} & larger values for $\sigma_8$, $\Omega_{\rm b} h^2$, and $\Omega_{\rm m} h^2$  \\     
    \hline
    \label{tab:models_other}
  \end{tabular}
\end{table*}

\subsection{Self-shielding using {\sc urchin}}
We post-process the \owls models using the {\sc urchin} reverse ray tracing radiative transfer code \citep{Altay_13}. In this scheme, the optical depth, $\tau$, around each gas particle is sampled in $N_{\rm ray}$ directions using rays of physical length $l_{\rm ray}$.  The optical depth is then used to relate the photoionisation rate $\Gamma^{\rm shld}$ at the location of the particle, to its optically thin value, $\Gamma^{\rm shld} = \Gamma^{\rm thin}\,\exp(-\tau)$.  We then use this shielded photoionisation rate to update the neutral fraction of the SPH particle and iterate the procedure until convergence. Most particles are in regions which are optically thin, $\tau\ll 1$, or optically thick $\tau\gg 1$ and converge very quickly. The majority of the computational effort is thus focused on those few particles in the transition region making the calculation very efficient.  Values of $N_{\rm ray}=12$, and $l_{\rm ray}=100$ physical kpc lead to converged results for the $z=3$ \ion{H}{I} CDDF in the simulations considered in this work.

In the \owls snapshots, the temperature stored for gas particles on the polytropic star-forming equation of state is simply a measure of the imposed effective pressure. When calculating collisional ionization and recombination rates, we set the temperature of these particles to $T_{\rm WNM} = 10^4$~K. This temperature is typical of the warm neutral medium phase of the ISM but our results do not change if we use lower values. We use case A (B) recombination rates for particles with $\tau < (>)1$. The optically thin approximation used in the hydrodynamic simulation leads to artificial photo-heating by the UV background for self-shielded particles.   To compensate for this, we enforce a temperature ceiling of $T_{\rm shld} = 10^4$ K for those particles that become self-shielded, i.e., attain $\tau > 1$. 

Early studies of self-shielding from the UV background relied on density thresholds \cite[e.g.][]{Haehnelt_98} to account for the attenuated photo-ionization rates in dense gas.  While this can provide an approximation to full radiative transfer schemes such as {\sc urchin}, it results in the transition between optically thin and optically thick gas being too sharp.  This was shown explicitly in \cite{Rahmati_13_evo} (see their Fig. 2) in which they compare density thresholds of $\nH = 10^{-1}, 10^{-2}$ and $10^{-3}$ cm$^{-3}$ to full radiative transfer results.  In addition, \cite{Rahmati_13_evo} suggested a fitting formula for the relation between hydrogen number density and photo-ionization rate that can be used to better approximate the results of full radiative transfer. However in this work we rely soley on results derived from ray tracing \owls models with {\sc urchin}.

\subsection{Local Sources}
\label{sect:localsources}
At the redshift of interest in this paper ($z=3$) a mostly uniform ionizing UV background pervades the Universe.  This background results from the integrated emission of a large number of sources which can be considered point like on cosmological scales.  Each source has a proximity zone in which its local radiation field is stronger than the integrated radiation field from all other sources.  Because these point sources of radiation cluster in regions of high gas density, it is possible that these proximity regions are important in calculating the abundance of \ion{H}{I} absorbers. In fact, several authors \citep[e.g.][]{Schaye_06,MiraldaEscude_05} have put forward idealized analytic arguments to support this idea.  In addition, \cite{Rahmati_13_str} recently used the \owls reference model to examine the effects of local sources on \ion{H}{I} absorbers in a cosmological galaxy formation setting.  Both these  analytic and numerical works found that local sources do not play an important role for optically thin absorbers, i.e. the Lyman-$\alpha$ forest.  For column densities between $10^{17.2} < \NHI < 10^{21} {\rm cm^{-2}}$, \cite{Rahmati_13_str} found that the inclusion of local sources can lower the normalization of the \ion{H}{I} column density distribution function (defined in \S 3) by approximately 0.25 dex. at $z=3$.  Above this column density, the Jeans scale of the gas is no longer resolved in the simulations used, and the effects become less certain.  In calculating self-shielding from a UV background, the main uncertainties are simply the amplitude of the radiation field and the density field through which the radiation is passing.  Due to the necessarily approximate treatment of star formation in cosmological simulations, additional sources of uncertainty enter when modelling local sources.  In particular, the small scale structure of the ISM, the location and luminosity of the sources themselves, and the dissociation, if any, of molecular hydrogen \citep[see][for a discussion of these uncertainties]{Rahmati_13_str}.  In an effort to isolate the effects of \owls model variations, we have neglected the effects of point sources in this work.  However, it is possible that local sources of radiation will affect different \owls models in different ways.

\subsection{Molecular Hydrogen}
\label{sect:molecules}
We adopt a prescription to model molecular hydrogen based on observations by \cite{Blitz_06} of 14 local spiral galaxies to form an H$_2$ fraction-pressure relationship. Their sample includes various morphological types and spans a factor of five in mean metallicity, although the lowest metallicity of any galaxy in their sample is one fifth of solar. They obtain a power-law scaling of the molecular fraction, $R_{\rm mol} \equiv \Sigma_{\rm H_2}/\Sigma_{\rm HI}$, with the galactic mid-plane pressure, $R_{\rm mol} = \left( P_{\rm ext} / P_0 \right)^{\alpha}$ with $\alpha = 0.92$ and $P_0 / k_{\rm b} =  3.5 \times 10^4 \, {\rm K \, cm^{-3}}$.  Applying this relationship to particles on the \owls star forming equation of state to calculate a molecular mass fraction yields 
\begin{eqnarray}
\label{eqn:H2}
f_{\rm H_2} &\equiv& \frac{2 n_{\rm H_2}}{2 n_{\rm H_2} + \nH} 
 = [1 + A(\nH/\nH^{*})^{-\beta}]^{-1} 
\end{eqnarray}
where $A = (P^{*} / P_0)$, $P^*$ is the pressure at the star formation threshold density $\nH^*$ and $\beta = \alpha \gamma_{\rm eos}$.  This relationship i{\sc urchin}s used to remove H$_2$ from the density field before the radiative transfer calculation.  In what follows, models labeled ``with H$_2$'' or ``corrected for H$_2$'' are those in which we have considered the formation of H$_2$ and removed it from the density field, while models labeled ``without H$_2$'' or ``not corrected for H$_2$'' are those in which no hydrogen was allowed to become molecular.  Unless stated otherwise, an \owls model name refers to the calculation in which no H$_2$ correction was made.  

It is probable that the average metallicity of the local \cite{Blitz_06} sample is an upper limit to the metallicity of DLAs at $z=3$.  This would cause us to produce too much molecular hydrogen in our models.  The size of the effect can be seen in Fig. \ref{fig:ref_dla} in which models with and without H$_2$ corrections are shown.  We also note that any model with a larger pressure threshold $P_0$ would approximate a lower metallicity environment and produce a result between our extreme cases.  We will discuss this further in \S \ref{sect:dlas}.

\section{The Column Density Distribution Function}

The CDDF, $\fN$, is defined as the number, $n$, of absorption lines per unit column density, $d\NHI$, per unit absorption distance, $dX$,
\begin{equation}
\label{eq:fhi}
\fN \equiv \frac{d^2n}{d\NHI dX}\,.
\end{equation} 
Absorption distance $dX$ is related to redshift path $dz$ as $dX/dz = H_0 (1+z)^2 / H(z)$, where $H(z)$ is the Hubble parameter \citep{Bahcall_69}.  In this work we focus on the relatively rare systems with column densities above $\NHI = 10^{17.0} {\rm cm^{-2}}$, which we identify using the following procedure (see \citealt{Altay_11} for a full description).  Having calculated the neutral fraction $x\equiv \nHI / \nH$ for each SPH particle using {\sc urchin}, we project all gas particles along the $z$-axis onto a grid with $16,384^2$ pixels using Gaussian approximations to their SPH smoothing kernels.  This leads to column densities along hypothetical lines of sight with a transverse spacing of 381 proper $h^{-1}$ pc or about 3/4 the gravitational softening length at $z = 3$.  Each pixel (i.e., line of sight) is associated with an absorption distance equal to $\Delta X_1 = H_0 (1+z)^2 L_{\rm box} \, c^{-1}$ where $L_{\rm box}$ is the comoving box size and the total absorption distance is $16384^2 \times \Delta X_1$.  We then histogram the column densities and divide by the total absorption distance to compute $\fN$.  We have verified that our results are converged with respect to the projected grid resolution. In addition, to confirm that very rare alignments of dense systems along the line of sight did not introduce non-negligible errors, we verified that our results do not change if we partition the simulation volume into 32 slabs and project them independently.

We have increased the normalization of $\fN$ in all \owls models presented here by 0.25~dex to match the abundance of observed low column-density DLAs. This re-scaling allows for a more meaningful comparison between models (and observations) at high column densities.  This shift approximates the behaviour expected from using a lower UV background normalisation and a larger value for $\sigma_8$.  We are motivated by two facts.  Firstly, WMAP3 values for cosmological parameters were used in the \owls simulations in which $\sigma_8 = 0.74$.  More recent measurements from the WMAP and Planck \citep{Planck_13} satellites have found larger values with $\sigma_8 \ge 0.81$.   Secondly, we used the UV background model of \citealt{HM01} (HM01) in our radiative transfer post-processing.  However, observational determinations \citep{Bolton_07, Becker_07, FGCA_08, Becker_13} and newer theoretical models by the same group \citep{HM12} are consistent with at least a factor of two lower normalization.  We have previously shown \cite[see][Fig. 2]{Altay_11} that WMAP7 cosmological parameters combined with a lower UV background normalization can cause shifts in the normalization of $\fN$ of this magnitude. 

We would like to stress two things.  First, we are not using these arguments to claim that our models fit current data. Instead, we are attempting to show how our various model $\fN$ are related to high column density observations when constrained to match low column density (DLA)  observations.  Second, the majority of this work is concerned with comparing different \owls models and hence is not affected by a global shift in normalization.

\begin{table}
  \centering
  \caption{Percent contribution to $\fN$ at different column density thresholds in model {\sc ref} (rounded to the nearest multiple of 5 percent) due to gas sub-samples as reported by \protect \citealt{VanDeVoort_12} (see their Fig. 4).  Halo gas refers to gas inside a halo, but not in the ISM. In addition to the standard column density thresholds for LLSs and DLAs, we add a category called strong DLAs with a threshold of $\NHI = 10^{21.5} {\rm cm^{-2}}$. }
  \begin{tabular}{cccc}
    \hline
    \hline
    Absorber Type & LLS & DLA & Strong DLA \\
    Column Density [${\rm cm^{-2}}$] & $10^{17.2}$ & $10^{20.3}$ & $10^{21.5}$ \\
    \\
    Gas Sub-sample & \multicolumn{3}{c} {Percentage Contribution to $\fN$} \\
    \hline
    \hline
    $T_{\rm max} < 10^{5.5} {\rm K}$ & 95 & 90 & 75 \\
    $T_{\rm max} > 10^{5.5} {\rm K}$ & 5  & 10 & 25 \\
    \hline
    IGM      & 40 & 10 & 0  \\
    Halo Gas & 60 & 80 & 15 \\
    ISM      & 0  & 10 & 85 \\
    Future ISM & 45 & 70 & 15 \\
    \hline
    In Halo & 60 & 90 & 100 \\
    In Halo + Inflowing & 30 & 45 & 50 \\
    In Halo + Static & 20 & 30 & 35 \\
    In Halo + Outflowing & 10 & 15 & 15 \\
    \hline
    $M_{\rm halo} < 10^{11} {\rm M}_{\sun}$ & 50 & 65 & 25 \\
    $M_{\rm halo} > 10^{11} {\rm M}_{\sun}$ & 10 & 25 & 75 \\
    \hline
    \label{tab:fN_contributions}
  \end{tabular}
\end{table}

\subsection{Overview}

\begin{figure}
  \begin{center}
    \includegraphics[width=0.45\textwidth]{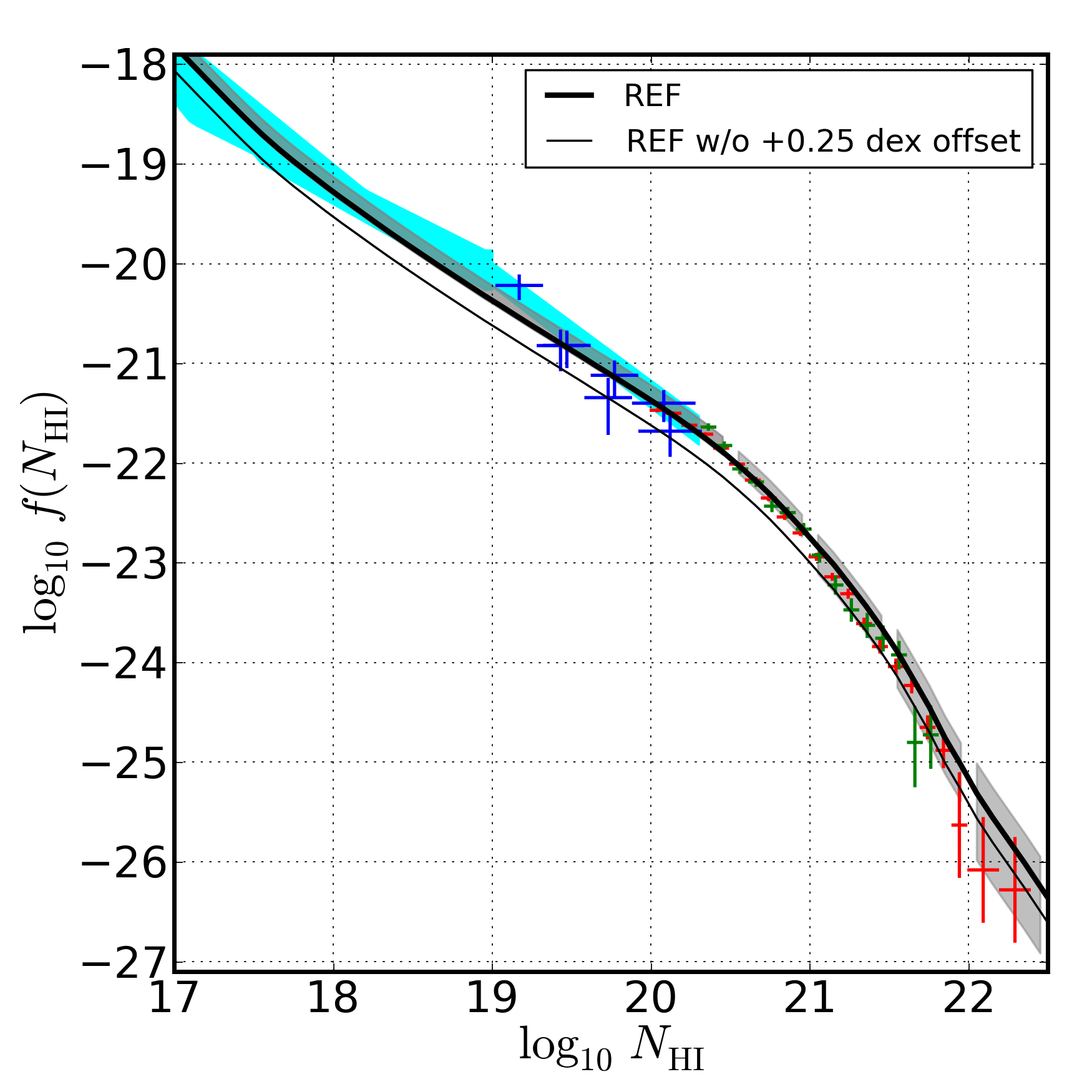}
  \end{center}
  \caption{ The $z=3$ \HI CDDF, $\fN$, in model {\sc ref}  compared to observational data from \protect \citealt{Omeara_07} (blue), \protect \citealt{Noterdaeme_12} (red), \protect \citealt{Prochaska_09} (green points with error bars) and \protect \citealt{Prochaska_10} (cyan shaded regions).   To account for more recent determinations of $\sigma_8$ and the amplitude of the UV background than were used in {\sc owls}, we also show model {\sc ref} with normalization increased by 0.25~dex (see the second paragraph of \S3 for more details).  The five grey bands are repeated from Fig. \ref{fig:main_cddf_big} and approximately indicate the variation between \owls models.  Molecular hydrogen was not allowed to form in these models.  }
\label{fig:ref}
\end{figure}

\begin{figure*}
  \begin{center}
    \includegraphics[width=0.99\textwidth]{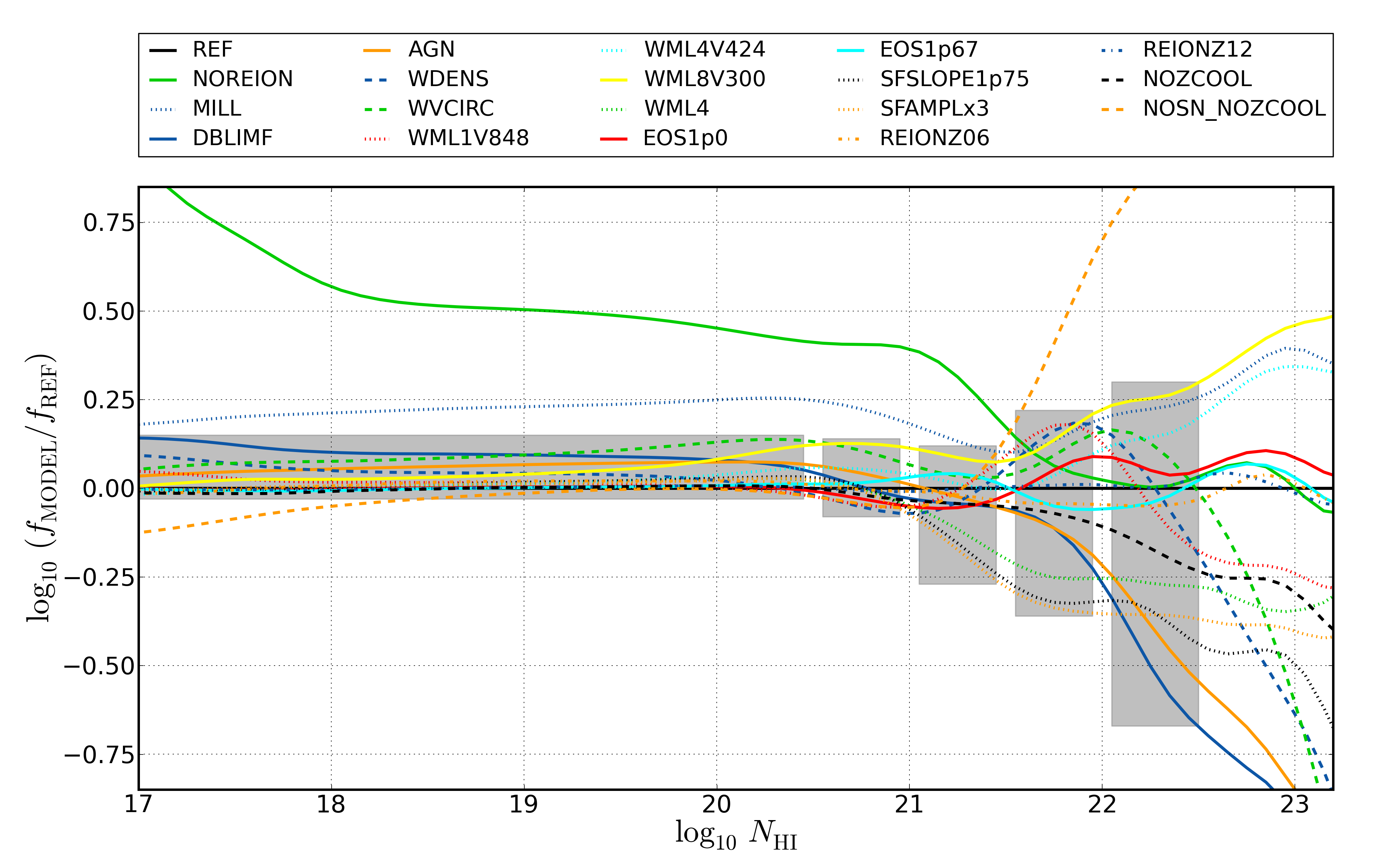}
  \end{center}
  \caption{Column density distribution functions in the \owls models relative to model {\sc ref} at $z=3$, $\log_{10} (f_{\rm MODEL} / f_{\rm REF})$. The leftmost grey box indicates a 0.2~dex range (a factor of 1.6) and bounds the variation between all \owls models in the LLS range excluding {\sc noreion}, {\sc mill} and {\sc nosn\_nozcool}.  Apart from these, the sub-grid physics implementation has only a very small effect on $\fN$ in the LLS range. The series of four grey boxes to the right bound the variation among the same \owls models in the DLA column density range.  The rightmost grey box extends to the highest column density for which observations are available.   Absorption lines in the strong DLA column density range probe the ISM and variations in $\fN$ due to sub-grid physics parameters become larger for larger $\NHI$. The inclusion of H$_2$ has a negligible effect for LLSs  but for DLAs it produces larger deviations from model {\sc ref} than shown here (see Figs. 8-15 below).}
\label{fig:main_cddf_big}
\end{figure*}

We start by presenting an overview of the results.  In Fig.~\ref{fig:ref} we show $\fN$ at $z=3$.  The thick and thin lines both show model {\sc ref}, but the thick line has a normalization adjustment as described above.  Observational constraints are taken from 
\cite{Omeara_07},
\cite{Prochaska_09}, 
\cite{Prochaska_10}, and
\cite{Noterdaeme_12}.
Model {\sc ref} captures the shape of $\fN$ as described by the latest observational data out to the highest measured column densities, $\NHI \sim 10^{22.4} {\rm cm^{-2}}$. 

In Fig. \ref{fig:main_cddf_big} we plot the $z=3$ \ion{H}{I} CDDF relative to model {\sc ref}. The five grey regions, which we also show in Fig. 1, illustrate the spread between physically realistic \owls models (i.e., excluding {\sc noreion}, {\sc nosn\_nozcool}, and also {\sc mill}).  The left-most region covers the entire LLS column density range while the right-most region covers the highest column densities for which observations exist. To provide the reader with a standard ruler of sorts, we repeat the five grey regions in every figure involving $\fN$.  Neither of these plots include models in which H$_2$ was allowed to form, but we will discuss H$_2$ formation in later sections.  Differences among physically plausible models in the LLS column density range are less than 0.2 dex indicating that the abundance of LLSs is robust to changes in sub-grid models.  However, the abundance of high column density DLAs is relatively sensitive to sub-grid model variations with the difference between models growing with $\NHI$, reaching approximately 1~dex at the highest observed column densities.   

The work of \cite{VanDeVoort_12} sheds light on these results and we briefly review it here.  They used the \owls {\sc REF} model to show (see Fig. 4 of their work) that more than 90\% of gas causing LLS absorption at $z=3$ has never been hotter than $10^{5.5}$ K.  Additionally, the contribution from gas in the ISM to $\fN$ is approximately 0\% at $\NHI = N_{\rm LLS} = 10^{17.2} {\rm cm^{-2}}$ and only 10\% at $\NHI = N_{\rm DLA} = 10^{20.3} {\rm cm^{-2}}$.  It was also shown that as column densities increase from $N_{\rm LLS}$ to $N_{\rm DLA}$, the fraction of absorption that is due to gas inside halos increases from 60\% to 90\%, and that the contribution from gas in halos that is either static or inflowing relative to the halo center goes from 50\% to 75\%.  In addition, the contribution from gas that is not in the ISM at $z=3$ but will be by $z=2$ (Future ISM) goes from 45\% to 70\%.  Consequently, we can characterise most of the LLS gas as fuel for star formation that is not in the smooth hot hydrostatic halo but rather in cooler denser accreting filaments. Conversely, the contribution of the ISM to DLAs goes from 10\% at $N_{\rm DLA}$ to 85\% at $\NHI = 10^{21.5} {\rm cm^{-2}}$ (a threshold column density we use to define ``strong'' DLAs). Therefore, most of the strong DLA gas traces the ISM and hence is affected by feedback from star formation.  For reference, we summarize these relationships in Table~\ref{tab:fN_contributions}.  

The majority of \owls variations examined in this work involve stellar feedback prescriptions.  If hot pressurized gas produced by SN explosions is able to escape the ISM, it tends to move away from star forming regions along the path of least resistance, avoiding overdense regions like filaments \citep[e.g.,][]{Theuns_02, Shen_13}.   This offers a simple explanation for the robust behaviour of LLSs with respect to changes in feedback prescriptions.  Much of the gas responsible for LLS absorption is not affected by feedback.  However, the LLS gas is not completely untouched by feedback and we do expect some changes.  For example, different feedback models will give rise to small changes in the interplay between hot pressurised outflows and cooler accreting components and will produce varying amounts of gas that is blown out by a wind only to cool and accrete back onto the galaxy.  

To conclude this overview, we note that 1) For all column densities that we examine, the majority contribution comes from gas in halos; 2) For these column densities, the \ion{H}{I} CDDF constrains the product of halo abundance and physical \ion{H}{I} cross-section per halo; 3) The halo mass function is very similar in all \owls\ models except {\sc mill}.  It follows that variations between models are driven by changes in the neutral hydrogen content of halos as a function of mass.  \cite{Haas_12_feed} and \cite{Haas_12_other} studied halo properties as a function of mass in the \owls models and we will make use of their results in what follows.

\subsection{Shape of $\fN$ in the LLS Regime}

\begin{figure}
  \begin{center}
    \includegraphics[width=0.5\textwidth]{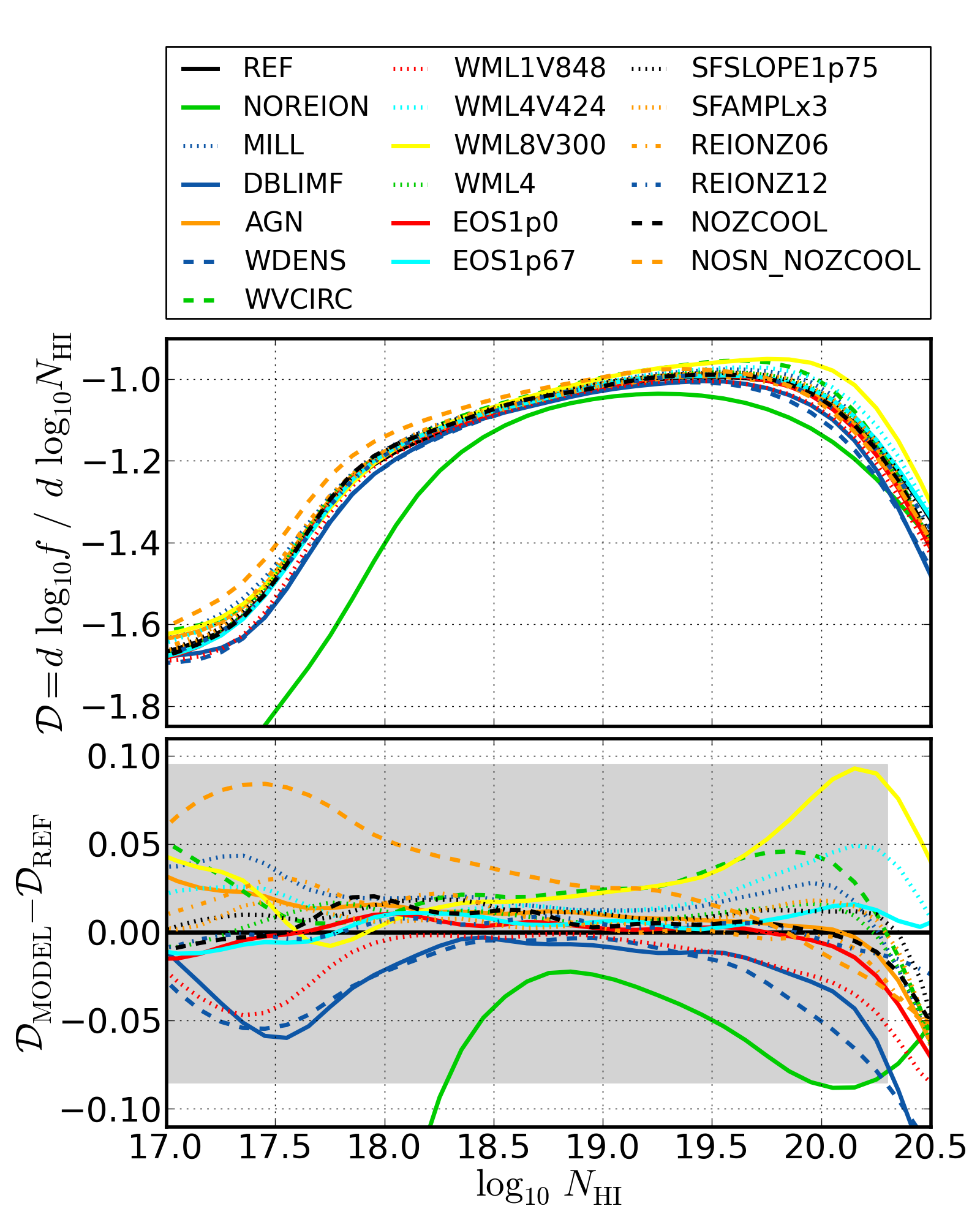}
  \end{center}
  \caption{ The slope of the $z=3$ column density distribution function in the LLS regime.   In the {\em upper panel} we show the derivative $\mathcal{D} = d \, \log_{10} f / d\, \log_{10} \NHI$ (i.e., $\mathcal{D}_{\sc REF}$ is the slope of the lines in Fig. \ref{fig:ref}).  In the {\em lower panel}  we show the difference with respect to model {\sc ref},  $\mathcal{D}_{\rm MODEL} - \mathcal{D}_{\rm REF}$.  The grey box has a height of 0.18 and bounds the variation in the LLS regime excluding model {\sc noreion}.  The shape of $\fN$ in this column density range is determined primarily by self-shielding from the UV background.   }
\label{fig:main_dfdn}
\end{figure}

In Fig.~\ref{fig:main_dfdn} we plot the logarithmic slope of $\fN$, $\mathcal{D} (\NHI) \equiv d \, \log_{10} f / d \log_{10} \NHI $, in the LLS column density range.  If $\fN$ were a power law, $\fN \propto N_{\rm H{\sc I}}^{-t}$, then $\mathcal{D} = -t$. All CDDFs were smoothed using a Hanning filter with FWHM of 0.4 decades in $\NHI$ before taking the derivative.  
In all \owls models which include a UV background (i.e., excluding {\sc no\_reion}), the CDDF has a characteristic shape.  At $\NHI = 10^{17} {\rm cm^{-2}}$ all models are consistent with a power law $\fN$ having slope $\mathcal{D} = -1.65$ with the spread among models approximately $\delta \mathcal{D} = \pm 0.05$.  The onset of self-shielding at $\NHI = 10^{17.2} {\rm cm^{-2}}$ causes $\fN$ to shallow.  The change in slope is initially rapid with $\Delta \mathcal{D} = 0.4$ between $10^{17} < \NHI / {\rm cm^{-2}} < 10^{18}$ and then proceeds gradually with $\Delta \mathcal{D} = 0.2$ between $10^{18} < \NHI / {\rm cm^{-2}} < 10^{20}$.  
Around the DLA threshold $\NHI = 10^{20.3} {\rm cm^{-2}}$ the neutral fraction saturates (i.e., the gas becomes fully neutral) which causes $\fN$ to steepen again.  The robust shape of $\fN$ among the \owls models indicates that self-shielding dictates the shape of the CDDF in the LLS regime.

\section{The Incidence of Absorption Systems}

\begin{figure*}
  \begin{center}
    \includegraphics[width=0.99\textwidth]{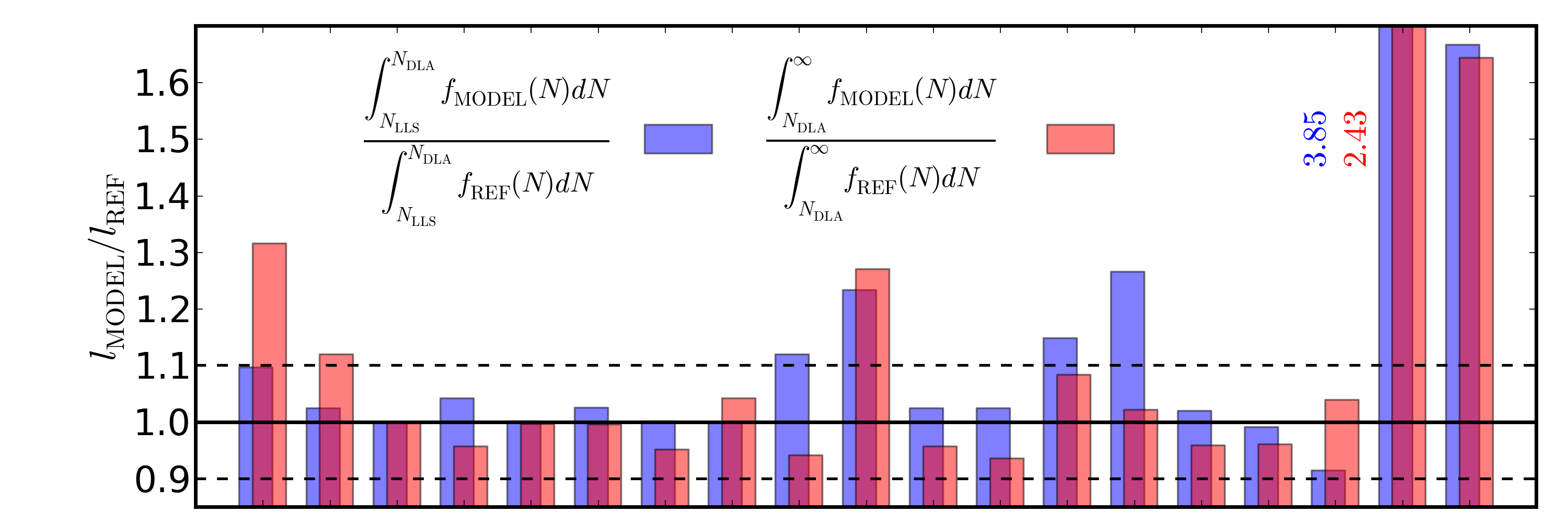}
    \includegraphics[width=0.99\textwidth]{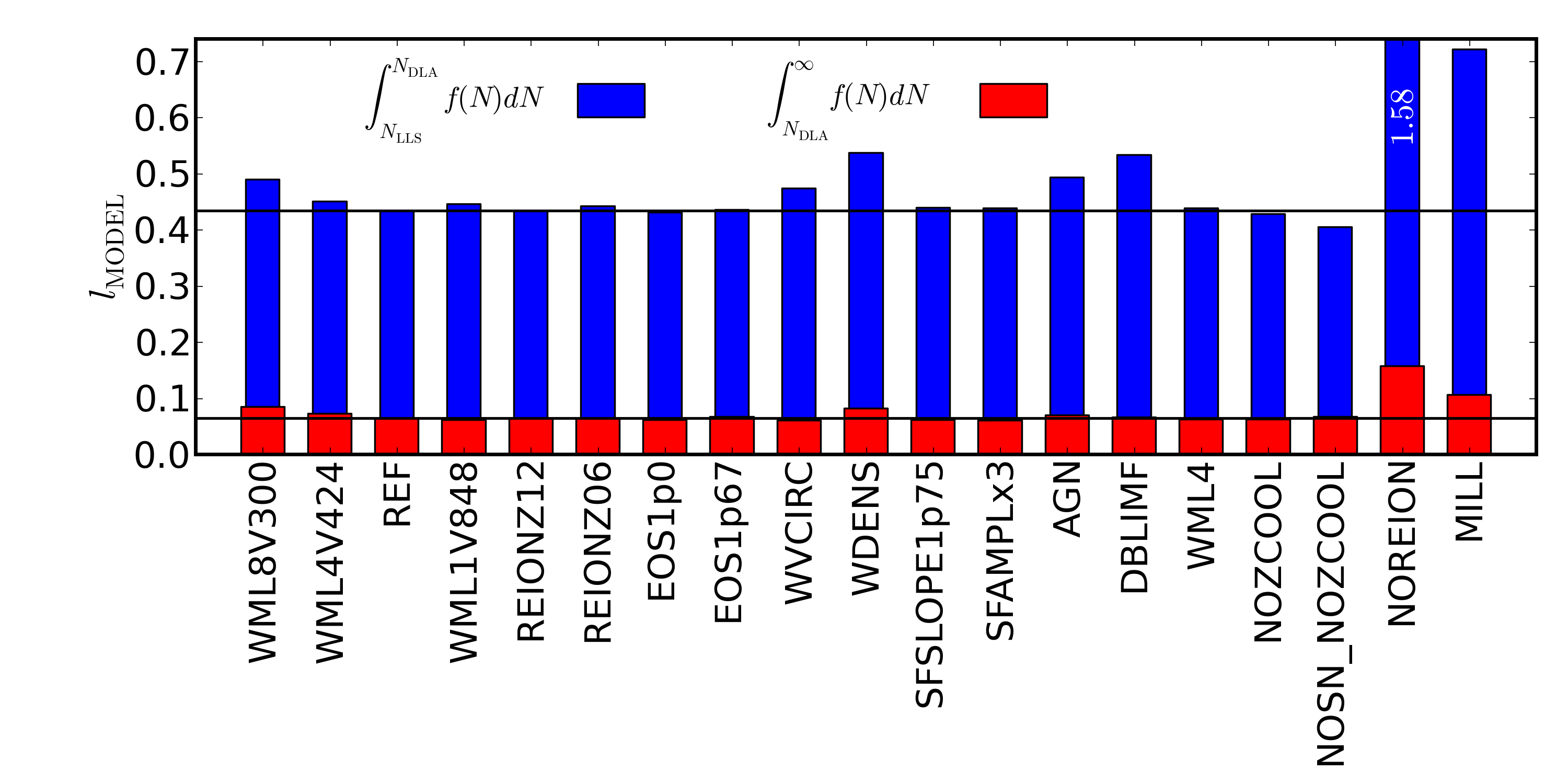}
  \end{center}
  \caption{The incidence, $l = \int_{N_{-}}^{N_{+}} f(N) dN = dn/dX$, of \ion{H}{I} absorbers per unit absorption distance at $z=3$.  The {\em bottom panel} shows $l_{\rm LLS}$ ($N_{-} = 10^{17.2} {\rm cm^{-2}}, N_{+} = 10^{20.3} {\rm cm^{-2}}$, blue) and $l_{\rm DLA}$ ($N_{-} = 10^{20.3} {\rm cm^{-2}}, N_{+} = \infty$, red).  The total height of each bar is equal to $l_{\rm LLS} + l_{\rm DLA}$.  Molecular hydrogen was not allowed to form in any of these models, but allowing H$_2$ formation has a negligible impact on $l_{\rm LLS}$ and $l_{\rm DLA}$ due to their insensitivity to strong DLAs which probe the ISM.  In the top panel we normalise $l_{\rm LLS}$ and $l_{\rm DLA}$ by the values in model {\sc ref}.   In both panels, solid horizontal lines indicate the values in {\sc ref}.  In the top panel, dashed lines indicate deviations of 10\% from {\sc REF}. Deviations from {\sc ref} in $l_{\rm LLS}$ and $l_{\rm DLA}$  are less than 10\% for most models.  Notable exceptions are {\sc mill}, which has a 60\% increase in both $l_{\rm LLS}$ and $l_{\rm DLA}$, driven mostly by the higher abundance of halos, and {\sc noreion} in which $l_{\rm LLS}$ and $l_{\rm DLA}$ are respectively factors of 3.9 and 2.4 larger due to the lack of Jeans smoothing and photo-evaporation.  The largest remaining variations follow trends in the halo gas fraction, $f_{\rm hg}$, as explained in the text.   }
\label{fig:inci}
\end{figure*}

The impact of sub-grid variations on the incidence of \ion{H}{I} absorbers, 
\begin{equation}
l = \int_{N_-}^{N_+} \fN dN_{\rm H{\sc I}} = \frac{dn}{dX}\,,
\label{eq:incidence}
\end{equation}
is shown in Fig.~\ref{fig:inci}. In the {\em bottom panel} we show $l_{\rm LLS}$ ($N_{-} = N_{\rm LLS} = 10^{17.2}$cm$^{-2}$,  $N_+=N_{\rm DLA} = 10^{20.3}$cm$^{-2}$) and $l_{\rm DLA}$ ($N_{-}=N_{\rm DLA}$, $N_{+}=\infty$) while in the {\em top panel} we normalise both results by the values in {\sc ref}.  Because $\fN$ has a logarithmic slope $\mathcal{D} \lesssim -1$, absorbers near $N_{-}$ make the dominant contribution to $l$.  This is evident in models such as {\sc nosn\_nozcool}, which deviates from model {\sc ref} by more than a factor of five at $\NHI = 10^{22.0} {\rm cm^{-2}}$ in Fig. \ref{fig:main_cddf_big}, but has incidences within 10\% of {\sc ref}.  Molecular hydrogen was not allowed to form in any of these models, but allowing H$_2$ formation has a negligible impact on $l_{\rm LLS}$ and $l_{\rm DLA}$ due to their insensitivity to strong DLAs which probe the ISM.

The magnitude of differences with respect to {\sc ref} are typically less than 10\% (dashed lines in top panel), but there are some notable outliers.  Model {\sc mill} was run with a higher value of $\sigma_8$ than {\sc ref} (0.9 as opposed to 0.74) and the same feedback parameters as {\sc wml4}.  It therefore contains more halos and more efficient feedback than model {\sc ref}.    The fact that the incidences $l_{\rm LLS}$ and $l_{\rm DLA}$ differ by less than 5\% 
between {\sc ref} and {\sc wml4} confirms that differences in cosmological parameters and not sub-grid physics make {\sc mill} an outlier.
Model {\sc noreion} did not contain an ionising background\footnote{ We did however perform radiative transfer on {\sc noreion} so the ionization state of gas was determined using the $z=3$ UV Background.  However, the gas was not subjected to any Jeans smoothing or heating from the UVB. }  
and therefore completely lacks any Jeans smoothing or photo-evaporation of gas from low-mass halos.  This increases the incidence of LLSs and DLAs by factors of 3.9 and 2.4 respectively. The fact that {\sc reionz12} and {\sc reionz6} have nearly the same CDDFs and incidences as {\sc ref} indicates that while Jeans smoothing and photo-evaporation are important, the abundance of \ion{H}{I} absorbers at $z=3$ is not sensitive to the timing of reionisation.  In addition, the convergence of models {\sc noreion} and {\sc ref} above $\NHI = 10^{21.5} {\rm cm^{-2}}$ indicates that strong DLAs are insensitive to the current amplitude of the UV background, as well as its history \citep[see also Fig. 2 in][]{Altay_11}.  We note here that part of this similarity at very high column densities may be due to the fact that we use a polytropic equation of state for the ISM as opposed to  explicitly simulating the physical processes that give rise to a multi-phase ISM.

Because the halo mass functions are very similar in all \owls models except {\sc mill}, differences in $l$ are due to differences in the gas distribution in and around halos of a given mass as opposed to the abundance of halos. 
In \cite{VanDeVoort_12} it was shown for model {\sc ref} that gas in low-mass ($10^9 < M_{\rm halo} / M_{\sun} < 10^{11}$) halos contributes 50\% to $\fN$ at $N_{\rm LLS}$ and 65\% at $N_{\rm DLA}$.  The remaining contributions to $\fN$ come primarily from the IGM at $N_{\rm LLS}$ (40\%) and from halos more massive than $10^{11} M_{\sun}$ at $N_{\rm DLA}$ (25\%).  The ISM of galaxies contributes negligibly at $N_{\rm LLS}$ and makes only a 10\% contribution at $N_{\rm DLA}$ (see Table \ref{tab:fN_contributions}).  While the exact magnitude of these contributions may vary between \owls models, we expect that in all models $\NHI = N_{\rm LLS}$ is associated with the transition from the IGM to halo gas and that $\NHI = N_{\rm DLA}$ probes the edge of the ISM.  We therefore expect that the halo gas fraction\footnote{The halo gas fraction is $f_{\rm hg} = (M_{\rm gas} - M_{\rm ISM}) / M_{\rm halo}$ and is called $f_{\rm gas,halo}$ in \cite{Haas_12_feed,Haas_12_other}}, $f_{\rm hg}$, in low-mass halos will be correlated with $l$ and that the correlation will be stronger for DLAs than for LLSs.

\cite{Haas_12_feed,Haas_12_other} examined the halo gas fraction (see panel E in their figures) as a function of total mass in the \owls models at $z=2$. While we do not expect $f_{\rm hg}$ in low-mass halos to be perfectly correlated with incidence, the trends in halo gas fraction for low-mass halos shown in \cite{Haas_12_feed,Haas_12_other} broadly mirror the variation of the DLA incidence seen in Fig.\ref{fig:inci}. 
\section{ Damped Lyman-$\alpha$ Systems - General }
\label{sect:dlas}

We find that, in the LLS column density range, $\fN$ is robust to changes in the sub-grid physics that is included.  However, the \owls variations, especially those dealing directly with star formation and feedback, can induce relatively large changes in the DLA column density range.  This is a natural consequence of the ISM making an increasingly large contribution to $\fN$ for higher column densities (\citealt{VanDeVoort_12} and Table \ref{tab:fN_contributions}).

\subsection{Galactic Outflows, Molecular Hydrogen, and Neutral Fraction Saturation Shaping $\fN$}

\begin{figure}
  \begin{center}
    \includegraphics[width=0.45\textwidth]{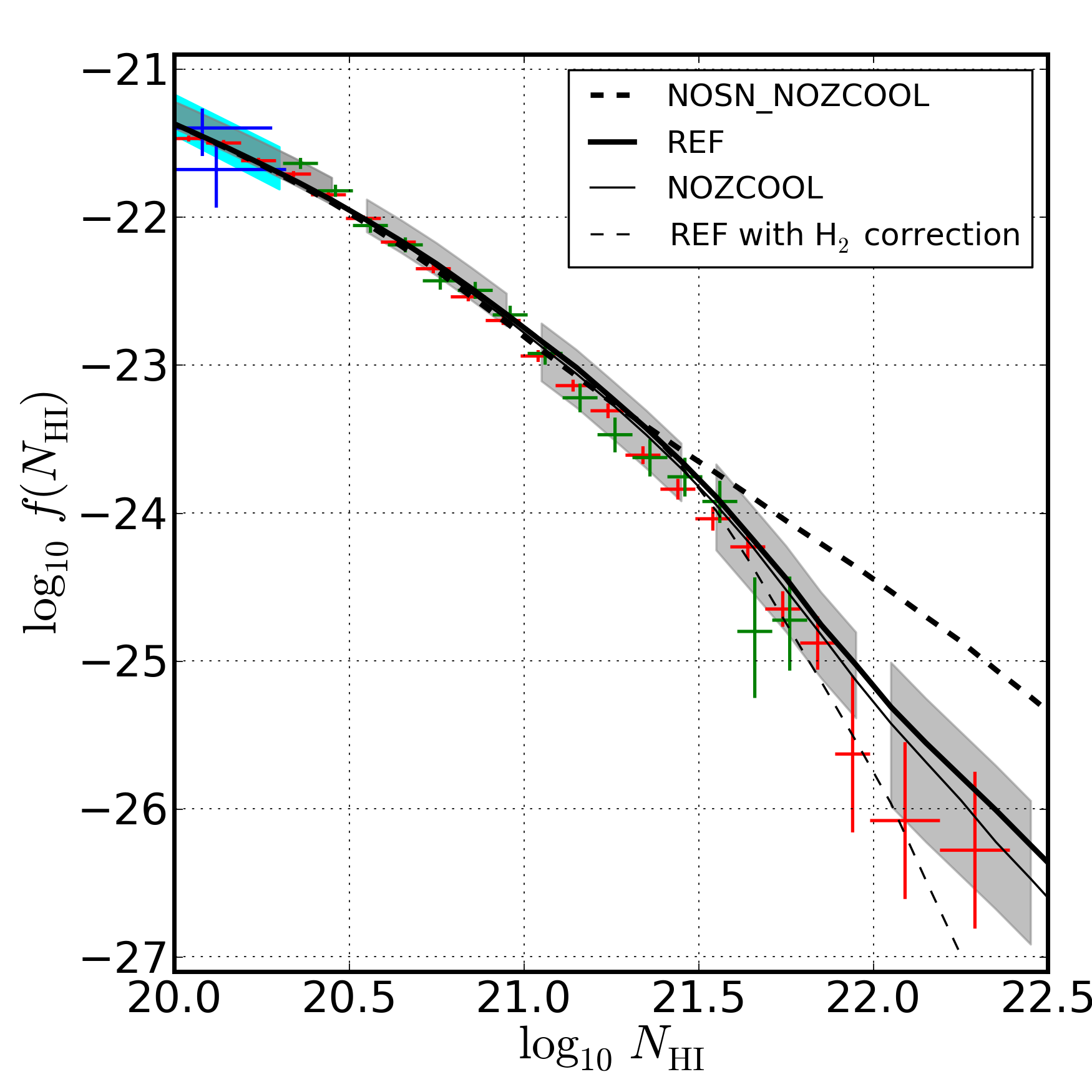}
  \end{center}
  \caption{The $z=3$ \HI column density distribution function $\fN$ in model {\sc nosn\_nozcool} ({\em thick dashed}), {\sc ref} ({\em thick solid}), {\sc nozcool } ({\em thin solid}), and {\sc ref}+H$_2$ ({\em thin dashed}).  Also shown is observational data from \protect \cite{Omeara_07} (blue), \protect \cite{Noterdaeme_12} (red), \protect \cite{Prochaska_09} (green points with error bars) and \protect \cite{Prochaska_10} (cyan shaded regions).  The five {\em grey regions} are the same as in Fig. \ref{fig:main_cddf_big} and approximately indicate the variation between \owls models (without H$_2$) in a given column density range.  Without feedback or H$_2$, $\fN$ is close to a single power law (model {\sc nosn\_nozcool}, dotted).  Models with either stellar feedback or molecular hydrogen have a break in $\fN$ between $10^{21.0} < \NHI / {\rm cm}^{-2} < 10^{21.5}$ and steepen thereafter.  The steepening due to feedback has the character of another power law while the steepening due to molecules has the character of an exponential cut-off.}
\label{fig:ref_dla}
\end{figure}

Before discussing the \owls variations in detail, we highlight three important physical processes: galactic outflows, molecular hydrogen formation, and saturation of the neutral fraction. In Fig. \ref{fig:ref_dla} we see that the model without feedback ({\sc nosn\_nozcool}) produces an $\fN$ that is approximately a power-law over the whole DLA column density range.  This behaviour is generic in simulations which lack sufficient feedback to drive winds and occurs independently of the form of feedback (thermal or kinetic) or of the choice of hydrodynamics solver (e.g., \citealt{Pontzen_08, Erkal_12, Bird_13}).  
Comparing models {\sc nosn\_nozcool} and {\sc nozcool}\footnote{The discussion of metal line cooling at this point is incidental.  Model {\sc nosn} could have been directly compared to {\sc ref} but was stopped before $z=3$.} allows us to isolate the effects of stellar feedback on $\fN$.  We see that both model {\sc nozcool} and {\sc ref} steepen relative to {\sc nosn\_nozcool}  between $10^{21.0} < \NHI / {\rm cm}^{-2} < 10^{22.0}$.  We will show shortly that every \owls model which includes feedback steepens in a similar way.  The SDSS catalogue of DLAs has been used to show that a single power-law cannot produce an acceptable fit to the observed CDDF data \citep{Prochaska_05, Noterdaeme_09, Noterdaeme_12}.  Efficient outflows generated by stellar feedback are a promising way to improve the agreement between model predictions and observations. 

In Fig.~\ref{fig:ref_dla} we also show model {\sc ref} with and without a correction for H$_2$.  Our implementation of molecular hydrogen also leads to a steepening between $10^{21.0} < \NHI / {\rm cm}^{-2} < 10^{22.0}$.  However, the steepening of $\fN$ due to H$_2$ is more severe than that due to outflows.  We note here that the position of the break due to H$_2$ should be treated as a lower limit.  The \cite{Blitz_06} relationship we used to correct for molecules applies to a local sample of galaxies which may have higher metallicities than the $z\sim 3$ galaxies associated with DLAs.  Employing a higher pressure threshold in Eq.~\ref{eqn:H2} would mimic lower metallicity environments and move the break from H$_2$ to higher column densities.  Next we will examine the shape of the CDDF more closely and show that the break due to H$_2$ formation is also generic among \owls models.

\begin{figure}
  \begin{center}
    \includegraphics[width=0.5\textwidth]{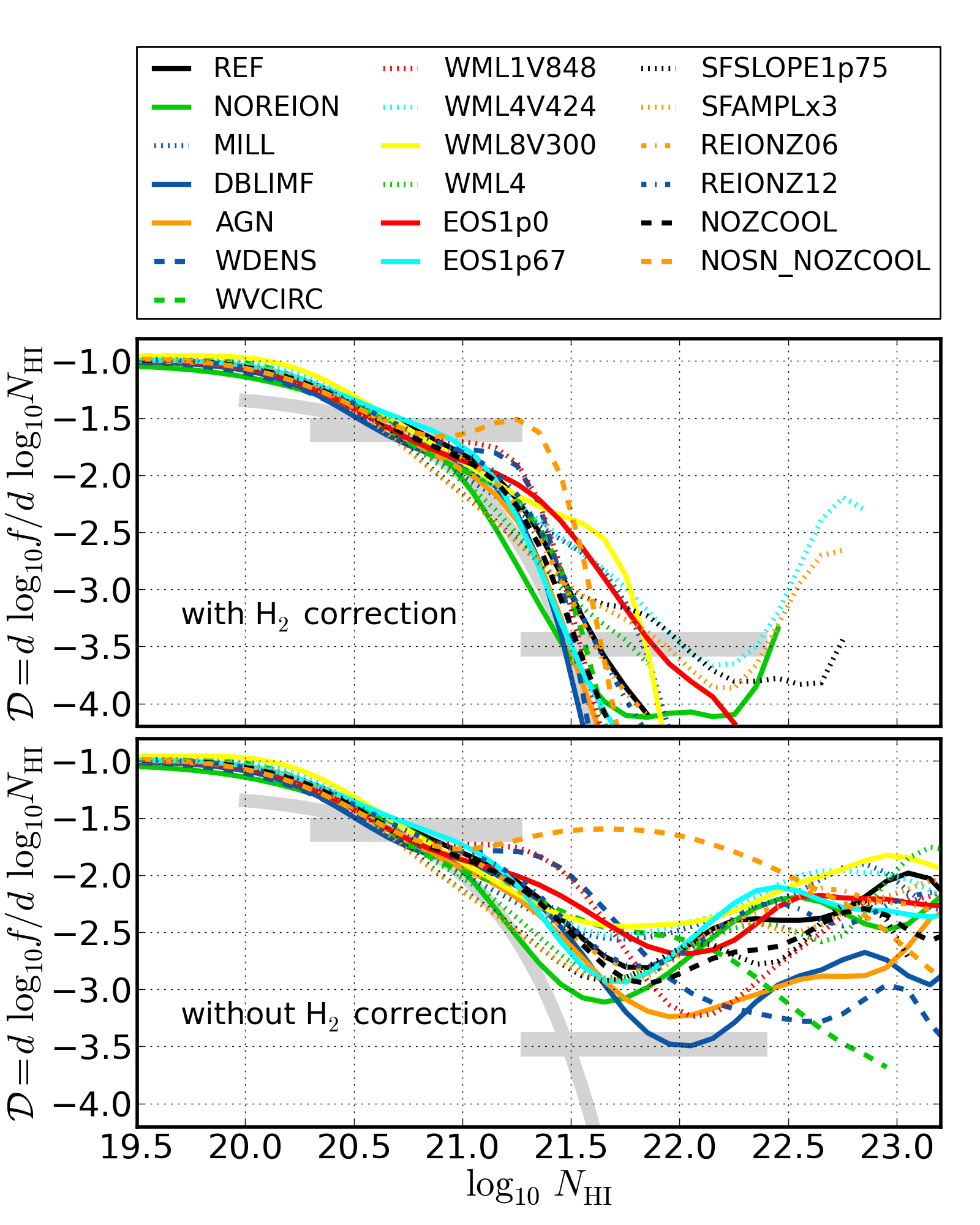}
  \end{center}
  \caption{The logarithmic slope of $\fN$, ${\mathcal D} \equiv d \, \log_{10} f / d\, \log_{10} \NHI$ for {\sc owls} models with ({\em top panel}) and without ({\em bottom panel}) a correction for molecular hydrogen. The two {\em horizontal grey regions} represent a double power-law fit to SDSS data by \protect \cite{Noterdaeme_09} extended to column densities observed in \protect \cite{Noterdaeme_12} while the {\em curved grey region} is a gamma function fit of the form $\fN \propto (N/N_{\rm g})^{\alpha_{\rm g}} \exp(-N/N_{\rm g})$ with $\alpha_{\rm g} = -1.27$ and $N_{\rm g} = 10^{21.26} {\rm cm^{-2}}$ by the same authors.  The steepening below $\NHI = 10^{21.0} {\rm cm^{-2}}$ from ${\mathcal D} = -1.0$ to ${\mathcal D} = -1.8$ is due to saturation of the neutral fraction around the DLA threshold and occurs in all models independent of the inclusion of feedback or H$_2$ correction.  At higher column densities and in models with high central pressures, ${\mathcal D}$  continues to decrease following closely the steepening characteristic of a gamma-function. Models without a correction for H$_2$ also continue to steepen for another decade in $\NHI$ due to galactic outflows (the one exception being the model without SN feedback {\sc nosn\_nozcool}).  These models then flatten or shallow slightly at even higher column densities.}
\label{fig:main_dfdnH2}
\end{figure}

In Fig.~\ref{fig:main_dfdnH2} we plot the logarithmic slope of the CDDF,  $\mathcal{D} \equiv d \, \log_{10} f / d\, \log_{10} \NHI$, and indicate the double power-law and gamma function fits to SDSS data by \cite{Noterdaeme_09} extended to column densities covered in \cite{Noterdaeme_12}.  The double power law appears as two horizontal grey regions while the gamma function is of the form $\fN \propto (N/N_{\rm g})^{\alpha_{\rm g}} \exp(-N/N_{\rm g})$ with $\alpha_{\rm g} = -1.27$ and $N_{\rm g} = 10^{21.26} {\rm cm^{-2}}$ and appears as a curved grey region.  We note here that both functional forms were acceptable fits to the data in \cite{Noterdaeme_09} but that only the double power-law produces an acceptable fit for the two data points above $\NHI = 10^{22} {\rm cm^{-2}}$ in \cite{Noterdaeme_12}.  These two points come from 5 (8) absorbers in the statistical (full) observational sample.  All {\sc owls} models have $\mathcal{D} \approx -1.0$ at $\NHI = 10^{19.5} {\rm cm^{-2}}$ and  steepen to $\mathcal{D} \approx -1.8$ at $\NHI = 10^{21.0} {\rm cm^{-2}}$.  This steepening is a self-shielding effect caused by the saturation of the hydrogen neutral fraction (i.e., the gas becoming fully neutral) around the DLA threshold and occurs in all \owls models (see also \citealt{Altay_11}).  In the model that does not include outlfows or an H$_2$ correction ({\sc nosn\_nozcool} in the {\em bottom panel} of Fig. \ref{fig:main_dfdnH2}) $\mathcal{D}$ is between -1.5 and -2.0 out to the highest observed column densities. 

Further steepening at higher column densities can be caused by outflows or the formation of molecular hydrogen.  We first concentrate on outflows by examining models in which we did not correct for H$_2$ (bottom panel of Fig. \ref{fig:main_dfdnH2}).  Above $\NHI = 10^{21.0} {\rm cm^{-2}}$, model {\sc nosn\_nozcool} (the only model that does not include outflows) flattens while all other models continue to steepen.  Models with low mass loading but high launch velocity ({\sc wml1v848} and {\sc wdens}) 
follow {\sc nosn\_nozcool} for approximately 0.25 dex in $\NHI$ before steepening.  These results indicate that the physical redistribution of gas due to efficient outflows is responsible for the steepening.  However even with efficient feedback,  models that do not correct for H$_2$ formation produce systems out to very high column densities of $\NHI = 10^{23.0} {\rm cm^{-2}}$. 

All models which correct for H$_2$ (top panel of Fig. \ref{fig:main_dfdnH2}) steepen further above $\NHI = 10^{21.0} {\rm cm^{-2}}$.  This steepening is more rapid than that which occurs due to outflows alone with many models reaching a slope as steep as $\mathcal{D} =-4.0$ between $10^{21.5} < \NHI / {\rm cm}^{-2} < 10^{22.0}$.  In contrast, models without a correction for H$_2$ steepen less, with most having $-3.0 < \mathcal{D} < -2.5$ in the same column density range.  Comparing the shape of \owls models to the gamma function fit from \cite{Noterdaeme_09} indicates that including a correction for H$_2$ introduces an exponential cut-off in $\fN$ in many models.  
This nearly universal cut-off due to molecules tends to homogenise the shape of $\fN$.  This is most evident for model {\sc nosn\_nozcool} which only flattens for approximately half a decade in $\NHI$ before being truncated by H$_2$ formation.  

Because the amount of molecular hydrogen included in our models should be considered an upper limit, models with and without H$_2$ corrections bracket the possibilities. \cite{Erkal_12} argued that the break around  $\NHI = 10^{21.0} {\rm cm^{-2}}$ in the observed CDDF is not caused by the formation of H$_2$. Their argument is based on the fact that the location of the break occurs at the same column density at $z=0$ as at $z=3$.  Gas at $z=0$ may have a higher metallicity and is exposed to a weaker UV flux than gas at $z=3$, and both effects would tend to favour a higher abundance of molecular hydrogen at lower $z$ for a given $\NHI$, hence  if the break were associated with H$_2$ one would expect the location of the break to shift to lower columns at $z=0$.  Our results are consistent with those of \cite{Erkal_12} in the sense that \owls models with and without an H$_2$ correction bracket the high column density slope of the \cite{Noterdaeme_09} double power-law.  This indicates that outflows can account for a large part of the steepening in the strong DLA column density range and that a higher pressure threshold (mimicking a lower metallicity environment) for H$_2$ formation would improve the agreement.
We conclude that while the displacement of high density gas by galactic outflows is an important model ingredient to reproduce the observed shape of the CDDF, metallicity dependent corrections for the formation of H$_2$ are important as well.



\subsection{Self-Regulated Star Formation and DLAs}

Previous analyses of galaxy properties in the \owls suite \citep{Schaye_10, Haas_12_feed, Haas_12_other} have demonstrated that star formation is self-regulated by the balance between accretion and galactic outflows.  Consider a typical star forming galaxy in which the accretion rate is determined by the host halo mass.  
If the current SFR does not produce enough feedback for outflows to balance the accretion rate, the ISM gas fraction, $f_{\rm ISM} = M_{\rm ISM} / M_{\rm halo}$, will increase. Given that the star formation rate is set by the gas surface density via the Kennicutt-Schmidt law \citep{Kennicut_98} in {\sc owls}, the star formation rate will increase in tandem, and with it the amount of feedback energy injected into the ISM.  Conversely, if the current SFR produces a level of feedback such that mass outflows exceed the accretion rate, $f_{\rm ISM}$ and the SFR will decrease, lowering the amount of feedback energy injected. Therefore, changes in the efficiency of star formation or feedback will lead to changes in $f_{\rm ISM}$ such that the amount of feedback that stars generate self-regulates. 

This effect of stellar feedback on the ISM gas fraction is particularly relevant for DLAs, since we have shown that high column density DLAs increasingly probe the ISM that feeds star formation. We expect that self-regulation will reduce the ISM gas fraction in the following models: 
({\em i}) less cooling ({\sc nozcool}) will decrease the accretion rate requiring less feedback to balance it, hence less star formation and thus less gas in the ISM;
({\em ii}) More efficient feedback ({\sc wml4}, {\sc agn}, {\sc dblimf}) requires {\em less} star formation for outflows to balance accretion, hence less gas in the ISM; 
({\em iii}) More efficient star formation or equivalently a shorter gas consumption time scale ({\sc sfamplx3}, {\sc sfslope1p75}) produces the same amount of feedback as {\sc ref} with less gas in the ISM and hence fewer high column density DLAs.

We note that only strong DLAs probe the star-forming ISM and so we do not expect these trends to be evident in statistics that are most sensitive to low column density DLAs such as the DLA incidence, but it may affect the cosmic density of \ion{H}{I} which is more sensitive to $f_{\rm ISM}$; we will discuss this and examine the DLA range of the \owls CDDFs in detail in the next section.

It is important to keep in mind that these expectations apply to the {\em total} amount of gas, whereas the CDDF only probes {\em neutral atomic} gas. Our model for molecule formation, described in Sect.~\ref{sect:molecules}, increases the molecular fraction of gas with increasing gas pressure. Variations of sub-grid parameters also lead to changes in the ISM's pressure which affects DLA statistics when H$_2$ corrections are included, in particular at column densities greater than  $10^{21.5}$~cm$^{-2}$. This can qualitatively alter DLA statistics, increasing, decreasing, or even reversing differences between models.

\subsection{Cosmological \ion{H}{I} density, $\Omega_{\rm HI,DLA}$}

\begin{figure*}
  \begin{center}
    \includegraphics[width=0.99\textwidth]{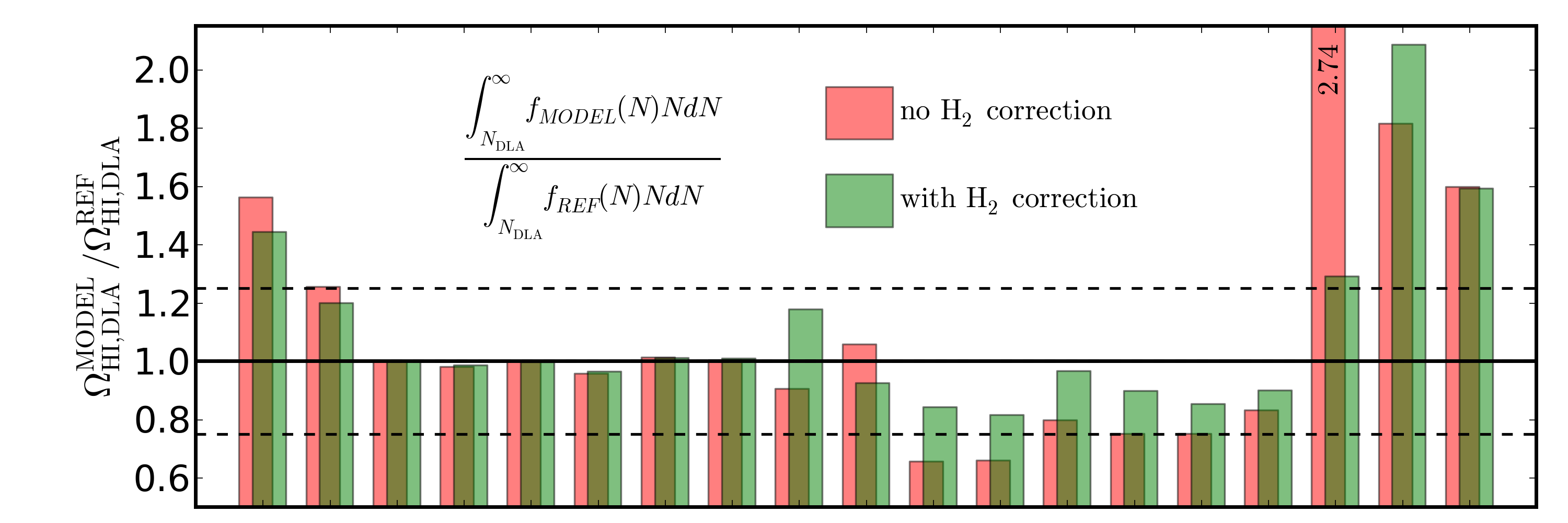}
    \includegraphics[width=0.99\textwidth]{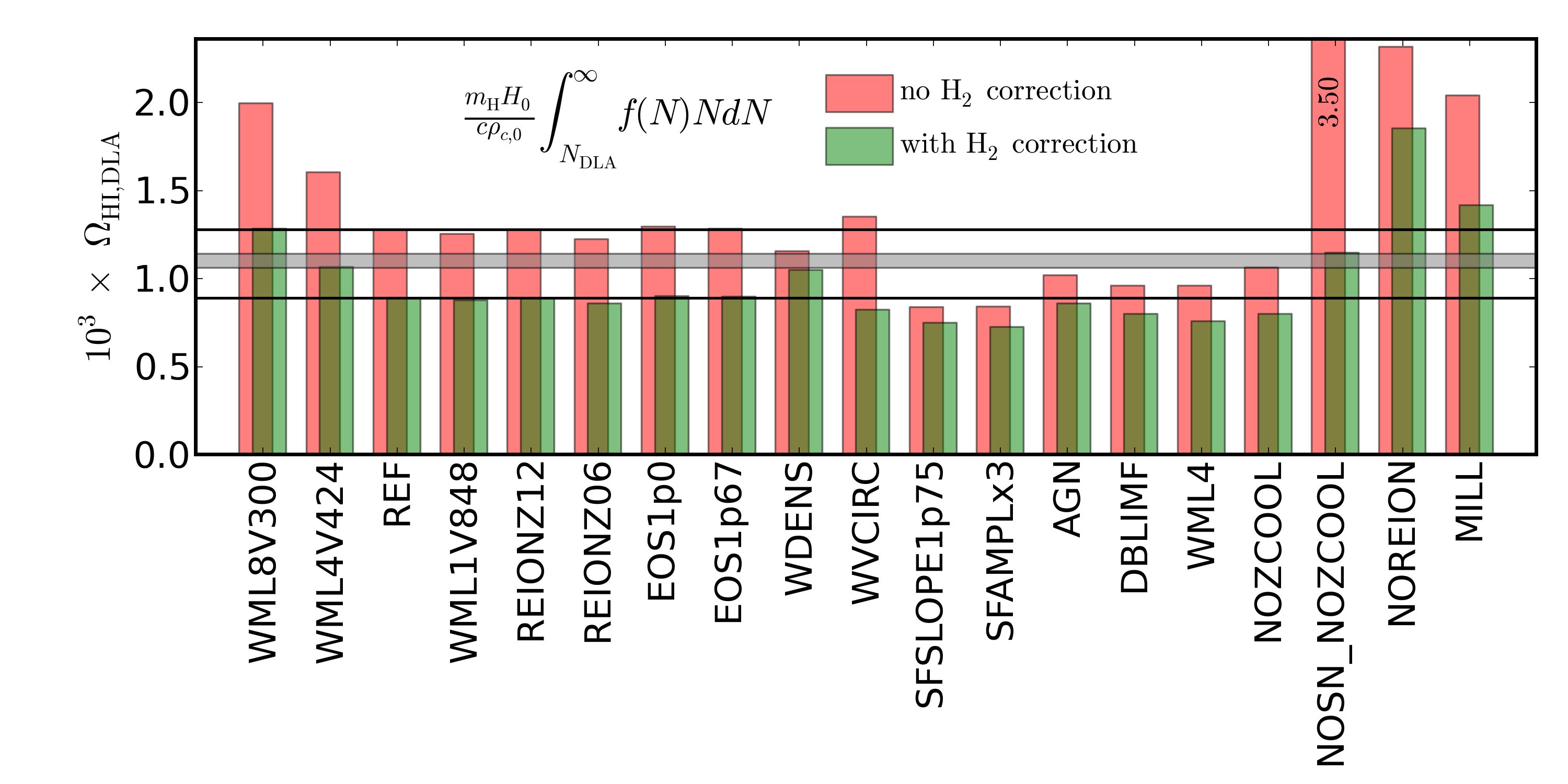}
  \end{center}
  \caption{ The cosmic density of neutral hydrogen in DLAs, $\Omega_{\rm HI,DLA}$ (Eq.\ref{eq:omega_HI}), in all \owls\ models at $z=3$ ({\em bottom panel}), and $\Omega_{\rm HI,DLA}$ for a given model compared to {\sc ref} ({\em top panel}); green and red bars represent models that have and have not been corrected for H$_2$, respectively. The {\em solid horizontal lines} in both panels indicate values in {\sc ref}, with the {\em dashed lines} in the top panel indicating deviations around {\sc ref} of 25 per cent.  The {\em filled grey band} in the bottom panel is the observational determination for $2.9 < z < 3.2$ from \protect \cite{Noterdaeme_12}.  Decreasing the efficiency of accretion, or increasing the efficiency of feedback or star formation, all reduce $\Omega_{\rm HI,DLA}$ due to self-regulation.  However, the differences are smaller if we correct for the presence of H$_2$.   
}
\label{fig:omega}
\end{figure*}

The cosmic abundance of neutral hydrogen in DLAs, $\Omega_{\rm HI,DLA}$,
\begin{equation}
\Omega_{\rm HI,DLA} = \frac{m_{\rm H} H_0}{c \rho_{c,0}} 
\int_{N_{\rm DLA}}^{\infty} \NHI \fN d\NHI,
\label{eq:omega_HI}
\end{equation}
where $\rho_{c,0}$ is the critical density at redshift $z=0$, is shown in Fig. \ref{fig:omega}.  This statistic is more sensitive to the high column density end of $\fN$ and hence to $f_{\rm ISM}$ than the incidence $l$ (see Fig.~\ref{fig:inci}).  Typical differences from {\sc ref} within the \owls\ suite are less than 25\% but there are some interesting trends.  Model {\sc mill} produces 60\% more \ion{H}{I} due to its higher number density of halos associated with the difference in cosmological parameters.  The surplus in model {\sc noreion} is smaller than in Fig.~\ref{fig:inci} due to the fact that the higher column density gas that contributes to $\Omega_{\rm HI,DLA}$ is not as sensitive to the evolution of the UV-background.   However, the surplus in model {\sc nosn\_nozcool} is much larger than in Fig.~\ref{fig:inci} due to the same increased sensitivity to high column densities.  The signature of self-regulation is visible in models {\sc nozcool}, {\sc wml4}, {\sc agn}, {\sc dblimf}, {\sc sfamplx3}, and {\sc sfslope1p75} as a reduced $\Omega_{\rm HI,DLA}$ compared to {\sc ref}. 
In addition, the trend with mass loading in models {\sc wml8v300}, {\sc wml4v424}, {\sc wml2v600} = {\sc ref}, and {\sc wml1v848} that was present in Fig.~\ref{fig:inci} is still present, but in this case it is due to $f_{\rm ISM}$ in halos with $M_{\rm halo} > 10^{11} {\rm M}_{\sun}$ as opposed to $f_{\rm hg}$ in halos with $M_{\rm halo} < 10^{11} {\rm M}_{\sun}$ (see the relative contributions in Table~\ref{tab:fN_contributions} and \citealt{Haas_12_feed}). 

Correcting for H$_2$ reduces the amount of \ion{H}{I} in a model, as is clear from the bottom panel of Fig. \ref{fig:omega}, but the magnitude of the effect varies between models.  In general, models with high gas fractions such as {\sc nosn\_nozcool} are affected more strongly than models with low gas fractions such as {\sc sfslope1p75} and {\sc sfamplx3}.  In this way, the inclusion of H$_2$ homogenises the CDDFs.  As a consequence, the spread in the green bars around {\sc ref} is smaller than the spread in the red bars.

\section{ DLA Model Variations in Detail }

In this section we examine in more detail the impact of \owls variations on the DLA range of the \ion{H}{I} CDDF. Figs.~8-15 all have the same layout. The {\em top panels} show $\fN$ for a specific subset of \owls models both with and without a correction for H$_2$.  The {\em lower panels} show the ratio of each model to {\sc ref}\footnote{Variations with H$_2$ corrections are normalized by {\sc ref} with an H$_2$ correction whereas variations without H$_2$ corrections are normalised by {\sc ref} without an H$_2$ correction.},  $\log_{10} (f_{\rm MODEL} / f_{\rm REF})$.  The grey regions indicate the variation between all physical \owls models when corrections for molecular hydrogen are not included (i.e., the same grey regions as in Fig. \ref{fig:main_cddf_big}).

\subsection{$\fN$ - Variations in Feedback or Cooling}

\subsubsection{ Supernova Feedback and Metal Line Cooling }
\label{sect:cooling}

\begin{figure}
  \begin{center}
    \includegraphics[width=0.5\textwidth]{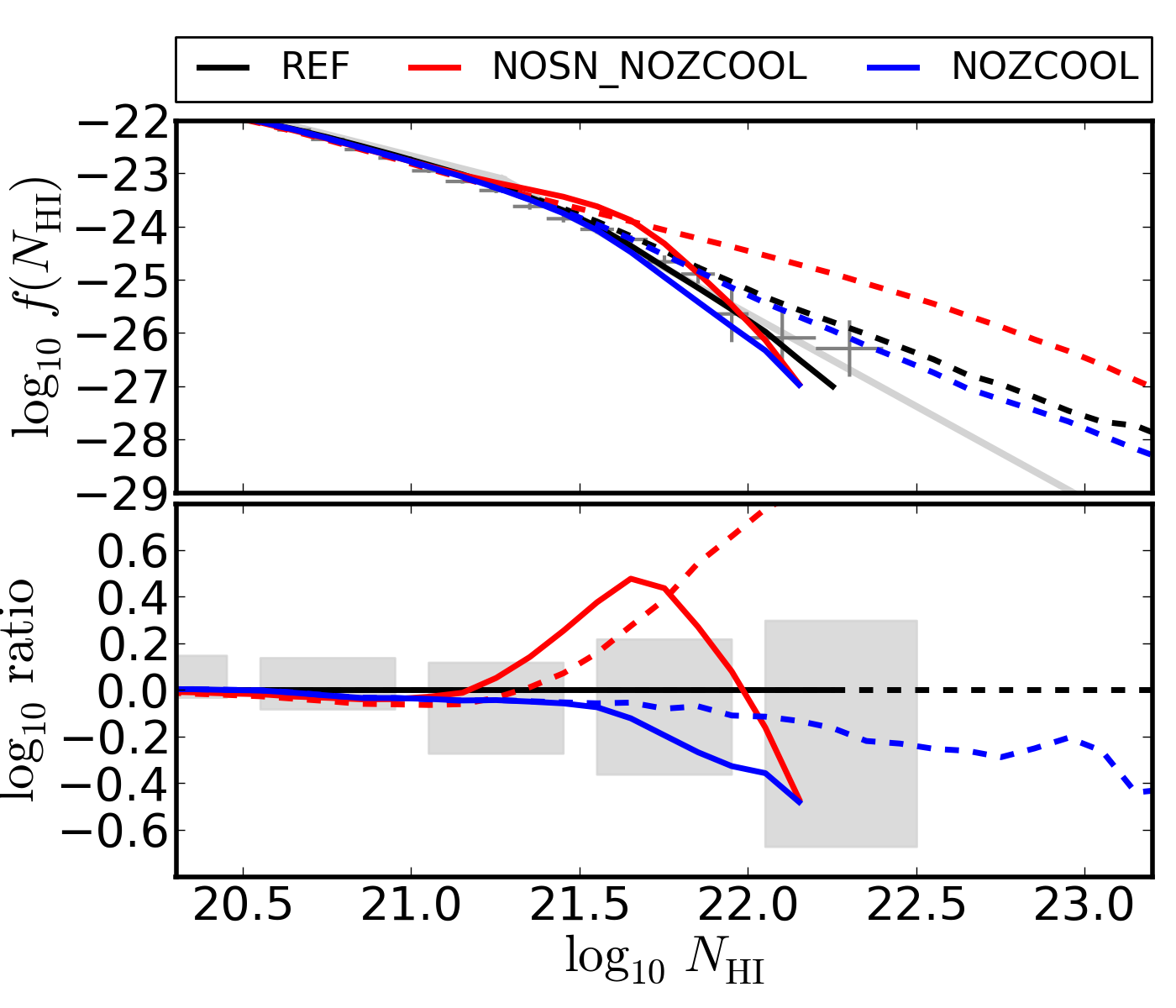}
  \end{center}
  \caption{The impact of SN feedback and metal-line cooling. \plotphrase.  
Shown in this plot is model {\sc ref} ({\em black}), a model with neither SN feedback nor metal-line cooling ({\sc nosn\_nozool}, {\em red}), and a model without metal-line cooling ({\sc nozcool}, {\em blue}). Neglecting metal-line cooling lowers the accretion rate onto galaxies which, due to self-regulation, produces a deficit in $\fN$ for strong DLAs.  Neglecting SN-driven winds increases the ISM gas fraction and hence leads to a surplus. 
However, when we correct for H$_2$ the large reservoir of ISM gas in {\sc nosn\_nozcool} becomes molecular and the trend reverses for $\NHI > 10^{22} {\rm cm^{-2}}$.
}
\label{fig:sn}
\end{figure}

The impact of SN feedback and metal-line cooling are shown in Fig. \ref{fig:sn}.  The lack of metal-line cooling in {\sc nozcool} decreases the accretion rate and hence, due to self-regulation, the ISM gas fraction, yielding a deficit with respect to {\sc ref}.  Including a correction for H$_2$ formation (solid curves) increases the difference between these models. Neglecting SN feedback allows gas to collect in the ISM producing nearly an order of magnitude more DLAs above $\NHI = 10^{22} {\rm cm^{-2}}$ than {\sc ref}.  However, when H$_2$ formation is accounted for, the higher ISM pressures in {\sc nosn\_nozcool} make molecule formation so effective that the abundance drops below {\sc ref} above $\NHI = 10^{22} {\rm cm^{-2}}$.   This is the clearest example of the inclusion of H$_2$ qualitatively changing the relationship of a model variation compared with {\sc ref}.

\subsubsection{Feedback Efficiency}
\label{sect:feedback}

\begin{figure}
  \begin{center}
    \includegraphics[width=0.5\textwidth]{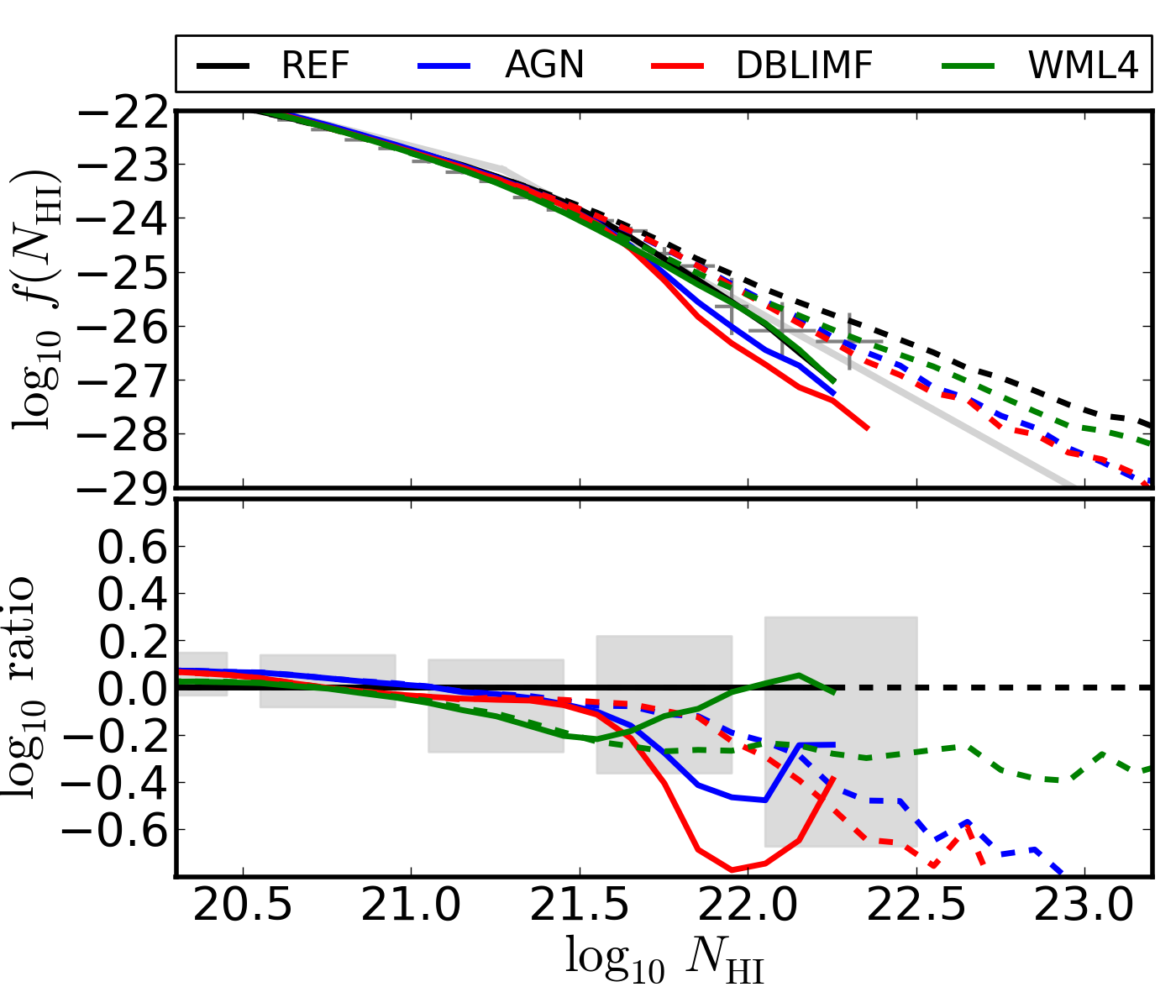}
  \end{center}
  \caption{As in Fig.~\ref{fig:sn}, but for the impact of strong feedback. We compare {\sc ref} to a model that includes AGN feedback ({\sc agn}, {\em blue}), a model with a top-heavy stellar IMF in star-bursts ({\sc dblimf}, {\em red}) and a model with twice the mass loading, $\eta$, of {\sc ref} ({\sc wml4}, {\em green}). More efficient feedback, due to self-regulation, produces deficits in $\fN$ with the difference becoming more pronounced at higher column densities. Including a correction for H$_2$ initially {\em increases} the differences with respect to {\sc ref}, but the sharp truncation due to H$_2$ homogenises models at even higher columns.}
\label{fig:agn}
\end{figure}

The impact of higher feedback efficiencies is shown in Fig.~\ref{fig:agn}.  Models {\sc agn} and {\sc dblimf}\footnote{Model {\sc dblimf} assumes a top-heavy IMF in high-pressure gas.  Although halos with a wide range of masses will contain high-pressure gas, this model will preferentially deposit energy into halos with deeper potential wells that contain more high pressure gas.}
inject extra feedback energy preferentially into high-mass halos while model {\sc wml4} injects twice as much feedback energy per stellar mass formed in all halos.  Models with more efficient feedback can balance a fixed accretion rate with a lower SFR and will have lower ISM gas fractions due to self-regulation. When the energy associated with such efficient feedback is injected democratically across all halo masses (model {\sc wml4}), as opposed to preferentially into high-mass halos (models {\sc agn} and {\sc dblimf}), the deficit of systems begins at a lower column density and does not become as large at higher column densities.  This is a consequence of the contribution of low-mass halos to $\fN$ becoming small at high column densities (see Table~\ref{tab:fN_contributions}). Correcting for molecule formation {\em increases} the differences with respect to {\sc ref} at low $\NHI$ but the models homogenise at high $\NHI$.

\subsubsection{ Mass Loading vs. Launch Velocity at Constant Energy }

\begin{figure}
  \begin{center}
    \includegraphics[width=0.5\textwidth]{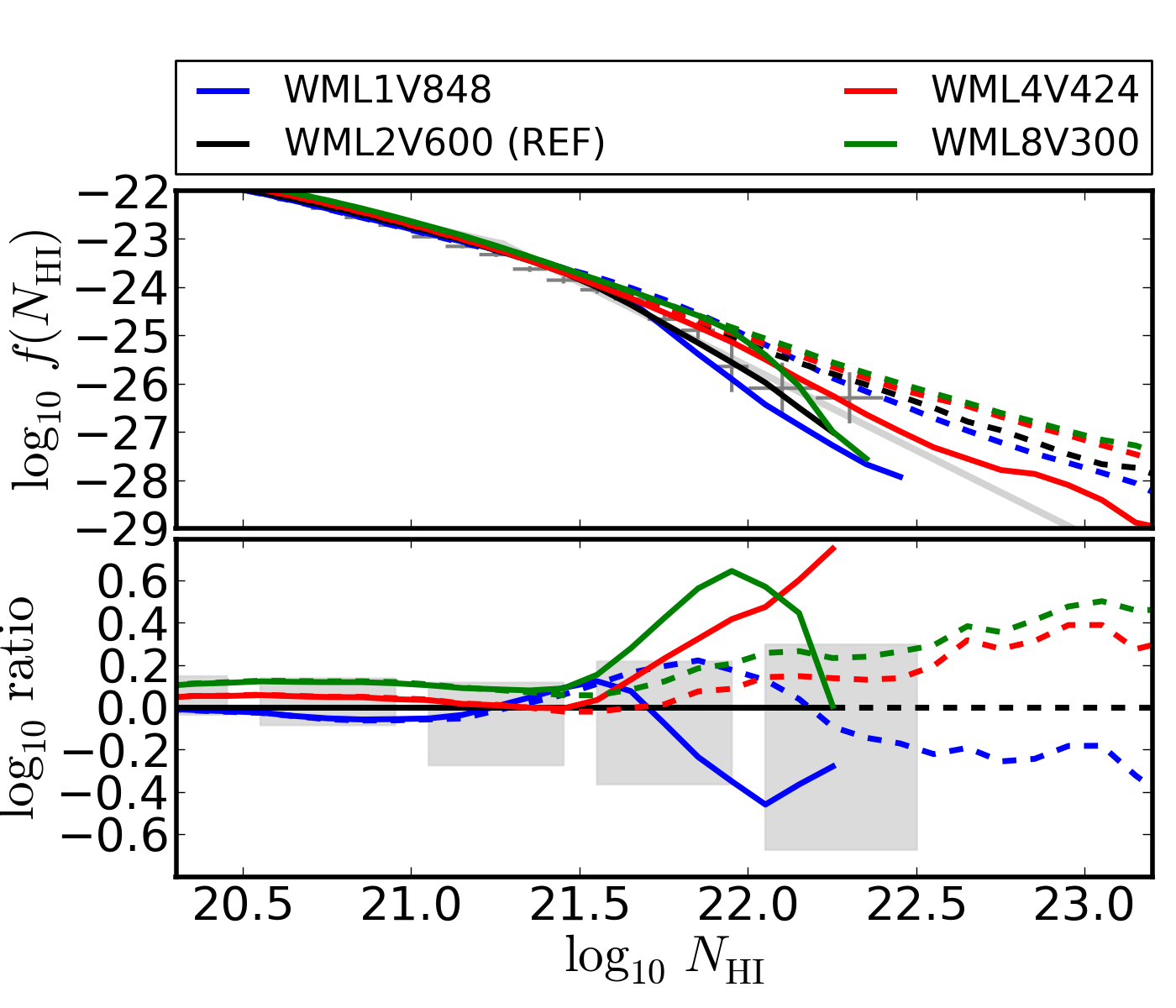}
  \end{center}
  \caption{As in Fig.~\ref{fig:sn}, but for the value of the launch velocity  $v_{\rm w}$ at constant energy injection per unit stellar mass, $\xi$. Lower  launch velocities result in more DLAs, especially above $\NHI = 10^{21} {\rm cm^{-2}}$  where the contribution from the ISM of more massive galaxies becomes important.  When we correct for H$_2$, the same trends are apparent at $\NHI < 10^{22} {\rm cm^{-2}}$ but truncation of $\fN$ due to H$_2$ formation obscures and eventually eliminates them as column density increases. }
\label{fig:wml}
\end{figure}

The efficiency with which SN feedback displaces gas is not solely dependent on the kinetic energy of the outflows, $\eta\,v_{\rm w}^2$. If the launch velocity is too low, 
gas will remain pressure confined in the ISM.   Because ISM pressure is correlated with halo mass, the depth of a halo's potential well sets up a threshold launch velocity below which SN feedback will not be effective at displacing gas from the ISM  \citep{DallaVecchia_08, DallaVecchia_12}.  When the quantity $\eta\,v_{\rm w}^2$ is fixed, outflows will be most effective in halos for which the launch velocity just exceeds the threshold. 

The trends in Fig.~\ref{fig:wml} can be explained by connecting the sub-selection of gas that makes the dominant contribution for a given column density (Table~\ref{tab:fN_contributions}) to the gas fraction trends with halo mass described in \cite{Haas_12_feed}.  \cite{VanDeVoort_12} showed that as column density increases the dominant contributor to $\fN$ in the DLA range is first halo gas in low-mass ($M_{\rm halo} < 10^{11} {\rm M}_{\sun}$) halos, then ISM gas in low-mass halos, then ISM gas in high-mass halos. \cite{Haas_12_feed} showed that: i) halo gas fractions in low-mass halos decrease as launch velocities increase, ii) ISM gas fractions in low-mass halos show evidence for an inversion of this trend, and iii) ISM gas fractions in high-mass halos again decrease as launch velocities increase (see panels D and E of Fig. 4 in their work).  The DLA abundances shown in Fig.~\ref{fig:wml} follow these trends in models without a correction for H$_2$. Including a correction for molecules enhances these trends before models are made similar by being truncated.  The only model in which $\fN$ is not truncated by H$_2$ formation is {\sc wml4v424} indicating that gas is able to reach high DLA column densities without exceeding the \cite{Blitz_06} pressure threshold.

\subsubsection{Environmentally Dependent Mass Loading and Launch Velocity }

\begin{figure}
  \begin{center}
    \includegraphics[width=0.5\textwidth]{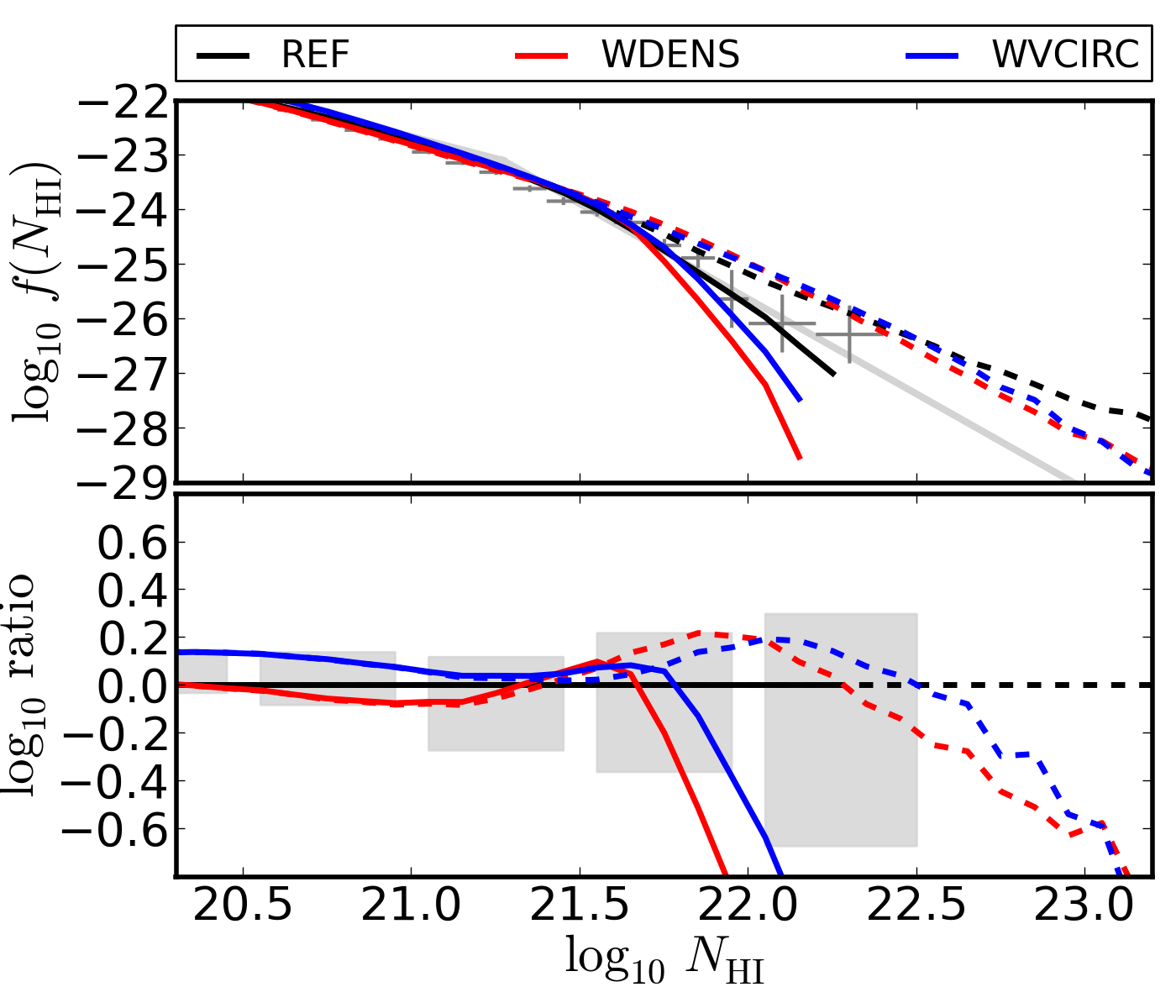}
  \end{center}
  \caption{As in Fig.~\ref{fig:sn} but for the dependence of wind speed on environment.  Along with model {\sc ref}, we show a model in which launch velocity increases with the local sound speed ({\sc wdens}, {\em red}) and a model in which launch velocity increases with the local velocity dispersion ({\sc wvcric}, {\em blue}). Compared to {\sc ref}, these models produce larger launch velocities in high-mass halos.  This results in a deficit of strong DLAs compared to {\sc ref}, particularly when we correct for H$_2$.}
\label{fig:whv}
\end{figure}

Models in which mass loading and launch velocity are dependent on environment are compared  in Fig.~\ref{fig:whv}.  The launch velocity in {\sc wdens} scales with the local sound speed, $v_{\rm w} \propto c_{\rm s,eos} \propto \nH^{1/6}$, and the mass loading is such that a constant amount of energy per stellar mass formed is injected, $\eta \propto v_{\rm w}^{-2}$.  The model is normalised such that the launch velocity is always equal to or greater than that of {\sc ref} and consequently the mass loading is always equal to or less than that of {\sc ref}.  
The launch velocity in model {\sc wvcirc} is proportional to the dark matter halo velocity dispersion, $v_{\rm w} \propto \sigma$, while the mass loading scales as $\eta \propto v_{\rm w}^{-1}$ in order to keep the momentum injected per unit stellar mass formed constant. 

Both models behave similarly to model {\sc wml1v848} for the reasons described in the previous section.  However, the increased launch velocity at higher densities ({\sc wdens}) and circular velocities ({\sc wvcirc}) produces more suppression at the highest column densities than {\sc wml1v848}.   This indicates that high velocity winds, even at very low {\em initial} mass loading, are effective at dragging along material that was not initially in the outflow.   Including corrections for H$_2$ in these models results in a dramatic decrease in the number of DLAs above $\NHI \sim 10^{21.5} {\rm cm^{-2}}$ indicating that these high velocity winds increase the halo mass corresponding to a fixed \ion{H}{I} column density, and hence the pressure in the ISM, which then turns molecular.

\subsection{$\fN$ - Variations Other than Feedback or Cooling}

\subsubsection{ ISM Equation of State }

\begin{figure}
  \begin{center}
    \includegraphics[width=0.5\textwidth]{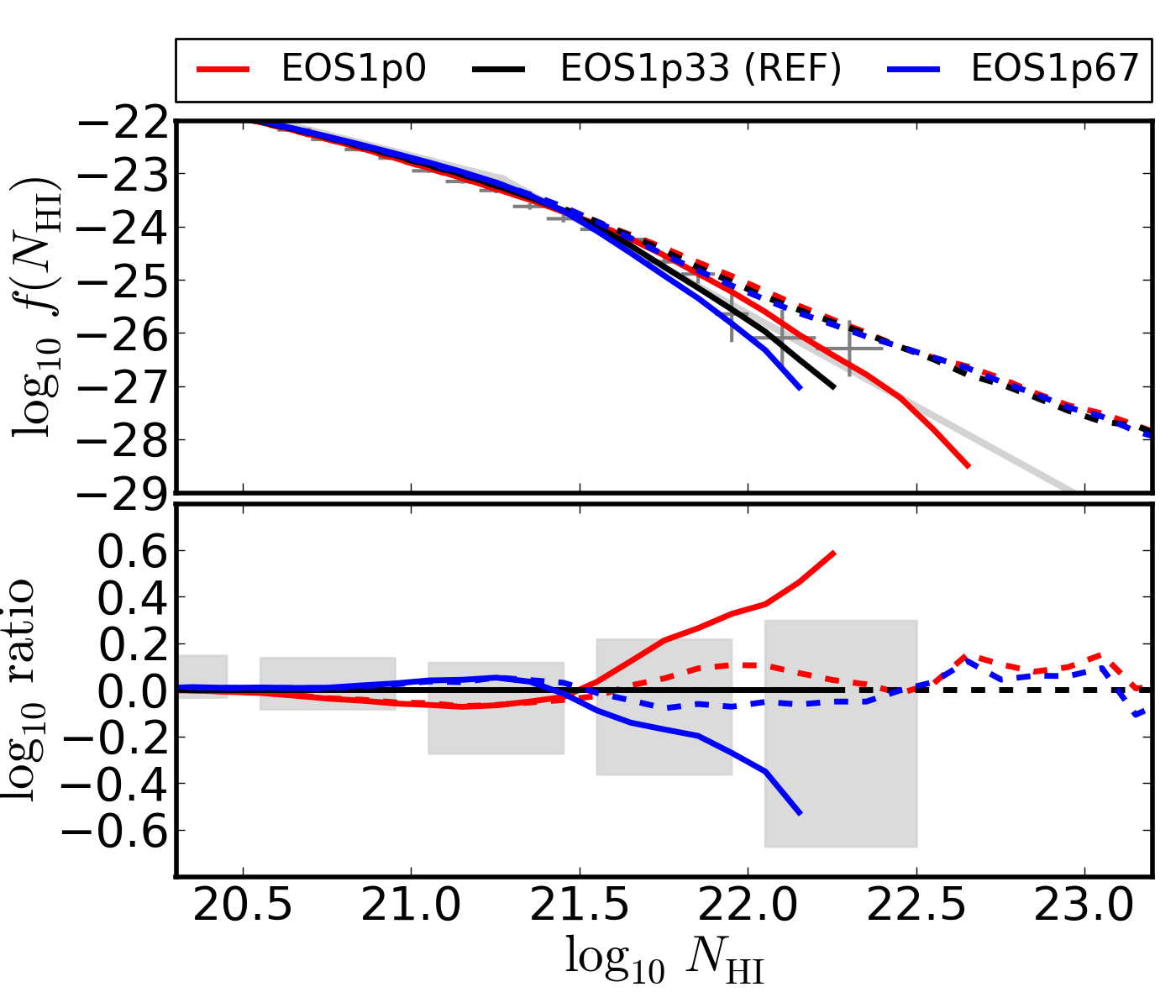}
  \end{center}
  \caption{As in Fig.~\ref{fig:sn}, but for the exponent in the equation of state for star forming gas, $p\propto \rho^{\gamma_{\rm eos}}$.  We compare models with $\gamma_{\rm eos}=1$ ({\em red}), 4/3 ({\sc ref}, {\em black}) and $\gamma_{\rm eos}=5/3$ ({\em blue}). Without correcting for H$_2$, the DLA abundance is not sensitive to the value of $\gamma_{\rm eos}$. However, when H$_2$ formation is accounted for, models with a stiffer equation of state (blue) have fewer strong DLAs than a more compressible model (red) due to the increased contribution of ISM gas to those column densities.}
\label{fig:eos}
\end{figure}

The impact of the choice of pressure-density relation in star forming gas, $p\propto \rho^{\gamma_{\rm eos}}$, is shown in Fig.~\ref{fig:eos}. Changes in the value of $\gamma_{\rm eos}$ affect the morphology of gas in disks \cite[e.g.,][]{Schaye_08,Haas_12_other}, but this has little effect on $\fN$. However, $\fN$ is sensitive to changes in $\gamma_{\rm eos}$ when a correction for H$_2$ is included, with more compressible gas producing significantly more DLAs above $\NHI = 10^{21.5} {\rm cm^{-2}}$ than a model with a stiffer equation of state.

\subsubsection{ Cosmological Parameters }

\begin{figure}
  \begin{center}
    \includegraphics[width=0.5\textwidth]{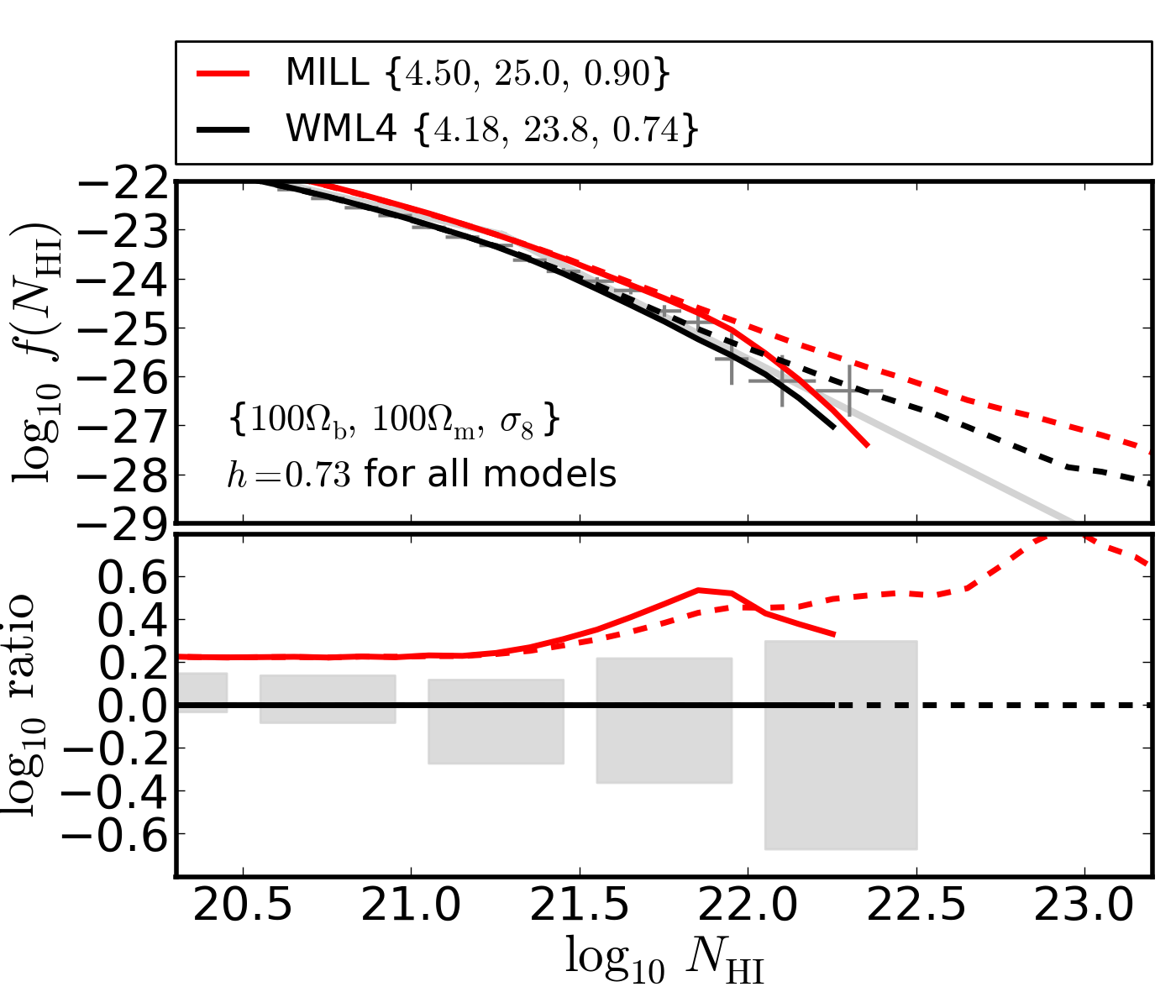}
  \end{center}
  \caption{As in Fig.~\ref{fig:sn}, but for cosmological parameters.  We compare model {\sc MILL}, which uses the Millennium cosmological parameters, with model {\sc wml4} that uses WMAP3 cosmological parameters. 
The greater density of matter and baryons, as well as the larger value for $\sigma_8$, in {\sc mill} lead to $\sim 0.2$~dex more DLAs nearly independent of column density up to $\NHI \le 10^{21.25} {\rm cm^{-2}}$, irrespective of the inclusion of H$_2$.  
At higher column densities {\sc mill} predicts a surplus of more than 0.2 dex. 
}
\label{fig:mill}
\end{figure}

Models {\sc mill} and {\sc wml4} differ only in terms of cosmological parameters and are shown in Fig.~\ref{fig:mill}.  Model {\sc mill} uses WMAP1 values  \{$\Omega_{\rm m}, \Omega_{\rm b}, \Omega_{\Lambda}, \sigma_8, n_s, h$\} = \{0.25, 0.045, 0.75, 0.9, 1, 0.73\}  whereas model {\sc wml4} uses WMAP3 values \{$\Omega_{\rm m}, \Omega_{\rm b}, \Omega_{\Lambda}, \sigma_8, n_s, h$\} = \{0.238, 0.0418, 0.762, 0.74, 0.951, 0.73\};
Therefore, model {\sc mill} has a higher physical matter density $\Omega_{\rm m} h^2$, physical baryon density $\Omega_{\rm b} h^2$, and linear amplitude of fluctuations $\sigma_8$.  

These differences produce a nearly constant surplus of approximately $0.2$~dex over the column-density range $10^{17} {\rm cm^{-2}} \le \NHI \le 10^{21.25} {\rm cm^{-2}}$, independent of the inclusion of an H$_2$ correction.  In fact, the ratio between the two models varies by only 2 per cent  between $\NHI = 10^{17.5} {\rm cm^{-2}}$ and $\NHI = 10^{21.0} {\rm cm^{-2}}$ (larger than range shown in plot). Above $\NHI = 10^{21.0} {\rm cm^{-2}}$ where the contribution from ISM gas becomes increasingly important, the difference between the two models increases further. 

The nearly constant offset between these models is driven by the different halo mass functions, suggesting it may be possible to use the abundance of DLAs to measure the growth of structure. As was shown in \cite{Altay_11}, changing the amplitude of the imposed UV background by factors of 3 and 1/3 produces a nearly constant offset of approximately 0.3~dex below the DLA threshold $\NHI = 10^{20.3} {\rm cm^{-2}}$. However, differences due to the UV background amplitude go smoothly to zero between $\NHI = 10^{20.3} {\rm cm^{-2}}$ and $\NHI = 10^{22.0} {\rm cm^{-2}}$.  This means there is likely an optimal column density around $\NHI = 10^{21.0} {\rm cm^{-2}}$ for isolating the effects of cosmological parameters from those due to the amplitude of the UV background.  Specifically, the effects from sub-grid variations have not become large at $\NHI=10^{21.0} {\rm cm^{-2}}$  (see Fig. \ref{fig:main_cddf_big}) and the low column density end of the DLA range is where $\fN$ is best constrained observationally.  Systems at this column density are readily identified at the resolution available in SDSS spectra and the abundance of systems is such that Poisson errors are smaller than 0.1~dex. 
However it must be kept in mind that local sources may effect the amplitude of $\fN$ at this column density \citep{Rahmati_13_str}.

\subsubsection{ Star Formation Law }
\label{sect:sfr}
\begin{figure}
  \begin{center}
    \includegraphics[width=0.5\textwidth]{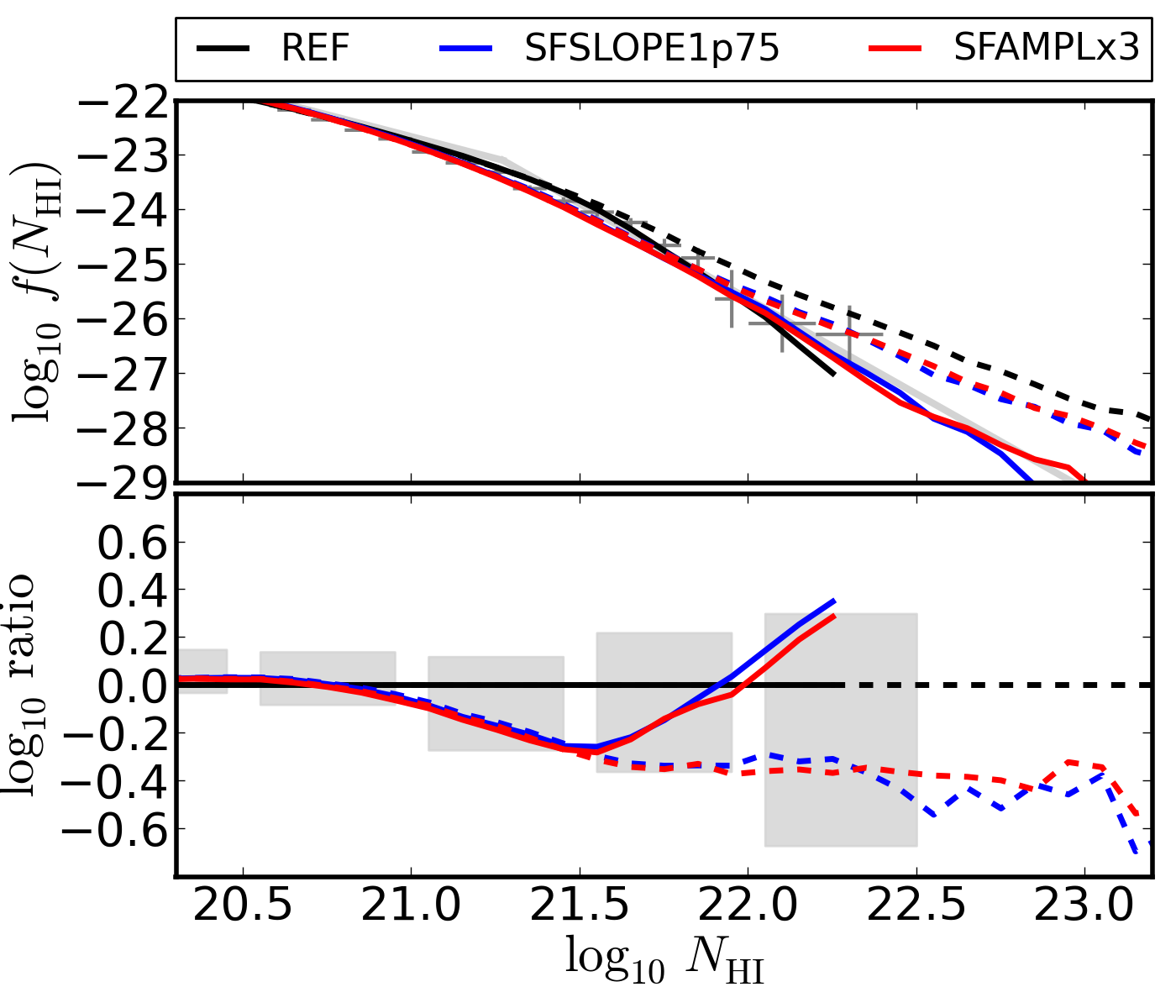}
  \end{center}
  \caption{ As in Fig.~\ref{fig:sn}, but for the assumed star formation law.  We compare {\sc ref} to a model in which the star formation rate is a steeper function of the gas surface density ({\sc sfslope1.75}, {\em blue}), and a model in which the star formation rate is a factor of 3 times higher for a fixed gas surface density ({\sc sfamplx3}, {\em red}).  Due to self-regulation, both model variations yield fewer strong DLAs than {\sc ref}.  When a correction for H$_2$ is added, the differences with respect to model {\sc ref} decrease above $\NHI = 10^{21.5} {\rm cm^{-2}}$, and become surpluses above $\NHI = 10^{22} {\rm cm^{-2}}$.
  }
\label{fig:sf}
\end{figure}

Changes in the assumed efficiency of star formation are shown in Fig.~\ref{fig:sf}.  Model {\sc ref} makes use of a Kennicut-Schmidt star formation law, $\dot\Sigma_{\rm SFR} = A_{\rm ks} \, (\Sigma_{\rm g}/M_\odot\,{\rm pc}^{-2})^{n_{\rm ks}}$  with $A_{\rm ks}=1.5\times 10^{-4}\,h^{-1}M_\odot\,{\rm yr}^{-1}\,{\rm kpc}^{-2}$ and $n_{\rm ks}=1.4$.  In model {\sc sfslope1.75}, the slope is increased to $n_{\rm ks}=1.7$, and in model {\sc sfamplx3} the amplitude, $A_{\rm ks}$, is increased by a factor of three.  These models, both of which assume more efficient star formation than {\sc ref}, are the cleanest demonstrations of self-regulation \citep{Schaye_10,Haas_12_other}.  In both cases, it takes less ISM compared to {\sc ref} to produce a SFR sufficient to drive outflows that balance a fixed accretion rate. The reduced ISM gas fractions translate directly into lower  abundances of strong DLAs.    

However, the trends are more complicated when a correction for H$_2$ formation is included.  The {\em lower} ISM gas fractions in galaxies with more efficient star formation in turn lead to lower central densities and pressures, and hence a smaller fraction of the gas becomes molecular compared to {\sc ref}. At column densities $\NHI \ge 10^{22} {\rm cm^{-2}}$, the latter effect dominates causing {\sc sfslope1p75} and {\sc sfamplx3} to produce more high column density DLAs than {\sc ref}.

\subsubsection{ Timing of Reionization }

\begin{figure}
  \begin{center}
    \includegraphics[width=0.5\textwidth]{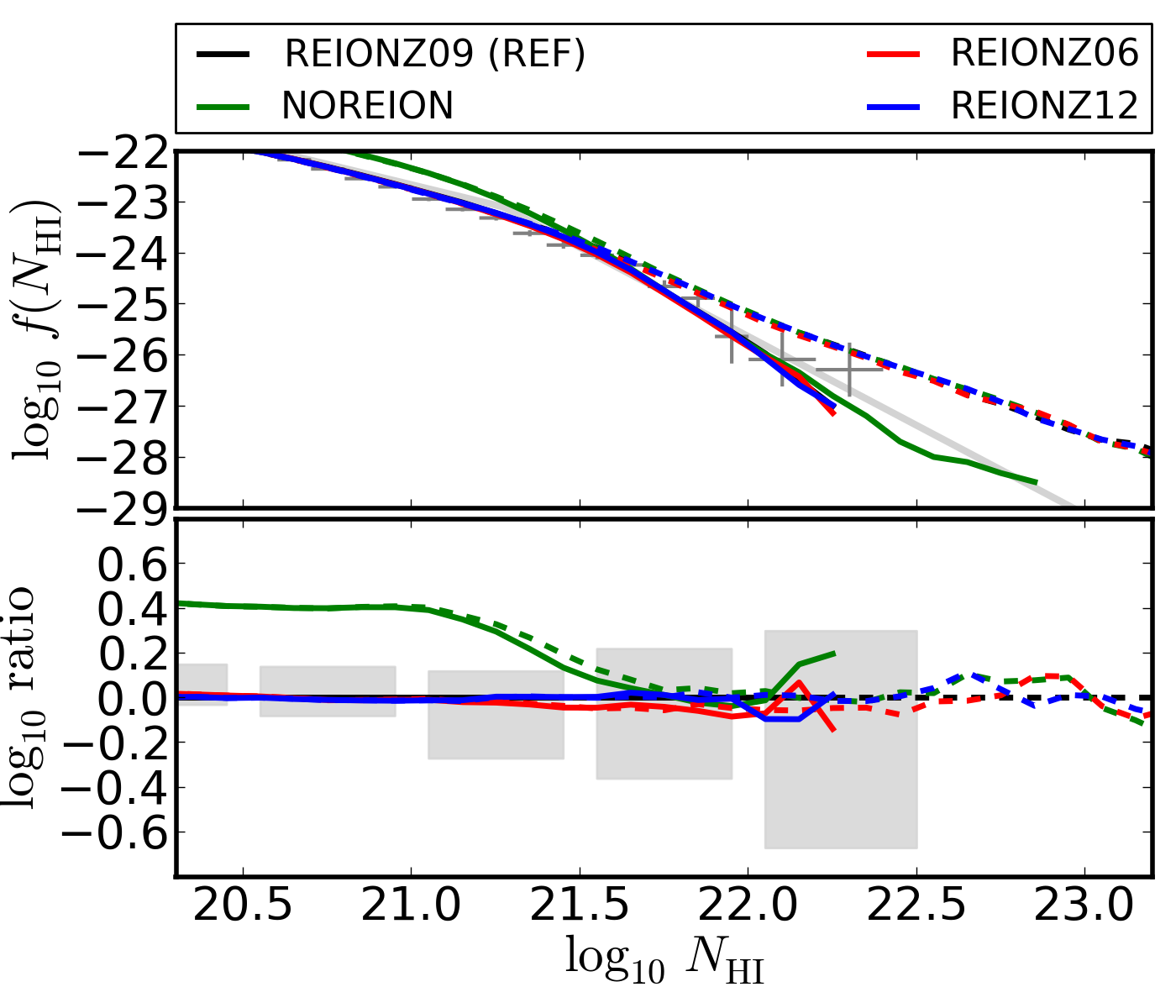}
  \end{center}
  \caption{ As in Fig.~\ref{fig:sn}, but for the timing of hydrogen reionisation.  We compare {\sc ref} ({\em black}), where reionisation occurred at $z_{\rm reion}=9$, to a model without reionisation ({\sc noreion}, {\em green}), and models with reionisation redshifts $z_{\rm reion}=6$ and $12$ ({\sc reionz06}, {\em red}) and ({\sc reionz12}, {\em blue}), respectively. Changing the redshift of reionisation over the range $[6,12]$ has no effect on the CDDF, but the model without reionisation has a constant offset from {\sc ref} until ISM (i.e., totally self-shielded) gas dominates $\fN$ at which point the results agree with {\sc ref}.  These trends are independent of the inclusion of H$_2$.    }
\label{fig:reion}
\end{figure}

Models with different reionization redshifts $z_{\rm reion}$ are shown in Fig.~\ref{fig:reion}.  The ionising UV background will quickly heat optically thin gas during reionisation. This will photo-evaporate gas out of sufficiently shallow potential wells, quench star formation in low-mass halos \citep[e.g.,][]{Okamoto_08}, make the gas distribution smoother than the underlying dark matter distribution \citep[e.g.,][]{Gnedin_98,Pawlik_09}, and induce peculiar velocities between gas and dark matter \citep[e.g.,][]{Bryan_99,Theuns_00}. By not including reionisation, the gas distribution in model {\sc noreion} is quite different from that in other \owls\, models.  As has been shown previously \cite[e.g.,][]{Theuns_02}, cosmic gas at $z=3$ retains very little memory of the history of reionization, provided it occurred sufficiently early.  The most important factor for DLA absorbers below $\NHI \approx 10^{21.5} {\rm cm^{-2}}$ is the current amplitude of the UV background.  At even higher column densities, the gas is totally self-shielded and is insensitive not only to the history of the UV background but also to its current amplitude.  The inclusion of molecular hydrogen does not significantly alter these model variations with respect to {\sc ref}. 
However, including an explicit treatment of the physical processes which give rise to a multi-phase ISM as opposed to using a polytropic equation of state may alter these conclusions.

\section{Discussion and Conclusions}




In this work we examined the effects of several physical processes on the  $z=3$ neutral hydrogen column density distribution function, $\fN$, by calculating this quantity in a set of 19 cosmological hydrodynamic simulations drawn from the \owls project \citep{Schaye_10}.  The project consists of a reference simulation ({\sc ref}), along with a large set of simulations that include systematic variations of the sub-grid model used in {\sc ref}.  We performed radiative transfer of the ionising UV background in post-processing to account for self-shielding using the code {\sc urchin} \citep{Altay_13}, and applied a phenomenological model for molecular hydrogen formation based on the results of \cite{Blitz_06}.   The treatment of H$_2$ relates the molecular hydrogen mass fraction in the ISM to gas pressure.  Because the star formation recipe in \owls is also based on gas pressure, the molecular hydrogen mass fraction and the star formation rate of the ISM gas are tightly related.

This work extends the high column density results of \cite{Altay_11} in which only model {\sc ref} was examined. In particular, we examined changes due to: {\em i})  the inclusion of metal-line cooling; {\em ii}) the efficiency of feedback from SNe and AGN, {\em iii}) the effective equation of state for the ISM; {\em iv}) cosmological parameters, {\em v}) the assumed star formation law and {\em vi}) the timing of hydrogen reionization.  To date, the \owls runs are the largest exploration of sub-grid parameter space relevant to cosmological galaxy formation simulations.  These kind of investigations are crucial to theories of galaxy formation as sub-grid effects are typically more important than the numerical method used to solve the equations of hydrodynamics \citep[see Appendix A and][]{Scannapieco_12}. 


In Fig. \ref{fig:ref} we compared $\fN$ in model {\sc ref} to observations.  The overall shape of our model $\fN$ agrees well with these observations but the model normalization is low by $\sim 0.25$~dex at the DLA threshold.  There are two contributing effects. First, the {\sc ref} model used WMAP3 values for cosmological parameters, $\{100 \Omega_m h^2, 100 \Omega_b h^2, \sigma_8\} = \{12.7,  2.23, 0.74\}$ which are all smaller than or very close to the WMAP7 values $\{13.4, 2.26, 0.81\}$ and the most recent Planck values $\{14.3, 2.21, 0.83\}$ \citep{Planck_13}.  Second, our radiative transfer calculation assumed the UV background normalization of \cite{HM01} but a more recent model by the same group \citep{HM12} has a factor $\sim 2$ smaller normalization at $z=3$ consistent with the most recent observational determination \citep{Becker_13}.  In \cite{Altay_11} it was shown that model {\sc ref} with WMAP7 cosmological parameters and a reduced UV background normalization comes very close to reproducing the observed abundance of absorbers.  However, the main goal of this work was not a comparison to observations but rather an investigation of the impact that sub-grid physics has on the column density distribution function (CDDF).  Therefore, we simply shifted the simulated CDDF by 0.25~dex when comparing models to data. This brings the models into agreement with observations in the low column density DLA range where observational error bars are smallest.  Our main conclusions are summarized below.

\subsection{Lyman Limit Systems}
The CDDF in the LLS column density range is robust to changes in sub-grid physics.  Comparing model variations to {\sc ref}, we find that the ratio $\log_{10} (f_{\rm MODEL} / f_{\rm REF})$ varies by less than 0.2~dex (a factor of 1.6) if we exclude models {\sc noreion} (which does not include a UV background), {\sc nosn\_nozcool} (which neglects stellar feedback), and {\sc mill} (which uses WMAP1 cosmological parameters).  We conclude that the abundance of LLSs is robust to changes in sub-grid physics but is sensitive to the assumed cosmology and the amplitude of the UV background.

The majority of LLS gas can be characterized as fuel for star formation that is not in the smooth hydrostatic hot halo but rather part of the colder denser accreting material \citep{VanDeVoort_12,Fumagalli_11}.  This material usually takes the form of filaments or streams.  Hot pressurized gas produced by SN feedback tends to move away from dense star forming regions along the path of least resistance.  This is why the abundance of LLSs does not change significantly in \owls variations that primarily affect the ISM or feedback strength  \cite[see also][]{Theuns_02}.

In Fig.~\ref{fig:main_dfdn} we showed that the shape of $\fN$, characterised by its logarithmic derivative $\mathcal{D} = d \, \log_{10} f / d\, \log_{10} \NHI$, takes on a universal form determined by self-shielding from the UV background. At $\NHI = 10^{17} {\rm cm^{-2}}$ all models are consistent with a power law $\fN$ with slope $\mathcal{D} \approx -1.65$ with the spread among models approximately $\delta \mathcal{D} = \pm 0.05$.  The onset of self-shielding at $\NHI = 10^{17.2} {\rm cm^{-2}}$ makes $\fN$ shallower.  The change in slope is initially rapid with $\mathcal{D}$ changing by 0.4 between $10^{17} < \NHI / {\rm cm^{-2}} < 10^{18}$ and then proceeds gradually with $\mathcal{D}$ changing by 0.2 between $10^{18} < \NHI / {\rm cm^{-2}} < 10^{20}$.  
Around the DLA threshold ($\NHI = 10^{20.3} {\rm cm^{-2}}$) the neutral fraction saturates (i.e., the gas becomes fully neutral) which causes $\fN$ to steepen again.

\subsection{Damped Lyman-$\alpha$ Absorbers}

\subsubsection{Saturation, Galactic Outflows, and Molecular Hydrogen}

We identified three physical processes that are important in shaping the HI CDDF: saturation of the neutral fraction, displacement of gas due to galactic outflows, and conversion of atomic hydrogen into molecules. The steepening of $\fN$ between $ 10^{20.3}<\NHI/{\rm cm^{-2}}<10^{21}$ is caused by saturation of the neutral fraction, $x \rightarrow 1$, and is independent of sub-grid model parameters and the inclusion of H$_2$.  Above this column density both galactic outflows and molecular hydrogen formation are capable of steepening $\fN$ further. Models in which feedback is inefficient or absent (e.g., \citealt{Pontzen_08,Erkal_12}, \owls model {\sc nosn\_nozcool}) and which have negligible molecular hydrogen corrections produce CDDFs that can be reasonably well characterized by a single power-law in the observed DLA column density range.  Once galactic outflows are introduced, the slope of $\fN$ continues to steepen up to $\NHI \approx 10^{22} {\rm cm^{-2}}$.  Current observations rule out a single power-law form for $\fN$ at high confidence \citep{Noterdaeme_12}.  

Correcting for the presence of molecular hydrogen based on the pressure threshold from \cite{Blitz_06} causes steepening of $\fN$ above $\NHI \approx 10^{21} {\rm cm^{-2}}$.  The shape of $\fN$ in this range matches the gamma function fit to observations made by \cite{Noterdaeme_09}, $\fN \propto (N/N_{\rm g})^{\alpha_{\rm g}} \exp(-N/N_{\rm g})$ with $\alpha_{\rm g} = -1.27$ and $N_{\rm g} = 10^{21.26} {\rm cm^{-2}}$, indicating that $\fN$ becomes exponentially suppressed in many of our models (see Fig.~\ref{fig:main_dfdnH2}).  \cite{Blitz_06} used a local sample of galaxies which likely have higher metallicities than $z=3$ DLAs \citep{Moller_13}. The column density above which \ion{H}{I} is converted into H$_{2}$ is predicted to increase with decreasing metallicity \citep{Schaye_01, Schaye_04, 2009ApJ...693..216K, 2010ApJ...714..287G}.  It would therefore be reasonable to expect the suppression of $\fN$ due to H$_2$ to occur at higher column densities than those we have shown here. In contrast, models without a correction for H$_2$ produce a shape more consistent with a double power-law.  Both exponential cut-off and  power-law relations are commonly used to fit the observed $\fN$, but the highest column density data points in \cite{Noterdaeme_12} would be difficult to accommodate with any fit that includes an exponential cut-off at  $\NHI \lessapprox 10^{22} {\rm cm^{-2}}$.  At present we conclude that a metallicity dependent cut-off due to H$_2$ formation as suggested by \cite{Schaye_01} is consistent with the data.



\subsubsection{Self Regulated Star Formation and DLAs}

\cite{Schaye_10}, \cite{Dave_11} and \cite{Haas_12_feed, Haas_12_other} have argued that star formation models which include feedback are largely self-regulating in the sense that changes in the efficiency of star formation, feedback, or accretion will lead to changes in the ISM gas fraction, $f_{\rm ISM}$, such that outflows driven by feedback balance gas accretion. 
This dependence on $f_{\rm ISM}$ is particularly relevant for DLAs and predicts the following relationships:

\begin{itemize}

\item{\bf Increased Feedback Efficiency.}
Models {\sc wml4}, {\sc agn}, and {\sc dblimf} all have increased feedback efficiency.  Therefore, a smaller $f_{\rm ISM}$ produces a SFR sufficient to drive outflows that balance a fixed accretion rate. 

\item{\bf Decreased Cooling Efficiency.}
Model {\sc nozcool} has less efficient cooling and therefore a reduced accretion rate for a given halo mass.  Therefore, a lower SFR and hence a smaller $f_{\rm ISM}$ is sufficient to produce outflows that balance the accretion rate. 

\item{\bf Increased Star Formation Efficiency.}
Models {\sc sfamplx3} and {\sc sfslope1p75} both have increased star formation efficiency and thus a shorter gas consumption time-scale.  Therefore, a smaller $f_{\rm ISM}$ is sufficient to produce a SFR that generates outflows capable of balancing the accretion rate. 

\end{itemize}


We indeed find these trends and show that they are most prominent in high column density DLAs for which the ISM contribution to $\fN$ is dominant (what we term strong DLAs). We explicitly showed that models with increased feedback efficiency ({\sc wml4}, {\sc agn}, {\sc dblimf}), decreased cooling efficiency ({\sc nozcool}), and increased star formation efficiency ({\sc sfamplx3}, {\sc sfslope1p75}) all produce deficits in $\fN$ with respect to {\sc ref} in the strong DLA range.  We note however that this reasoning only applies to the {\em total} amount of gas in the ISM, and not the {\em neutral atomic} hydrogen responsible for $\fN$. We have shown that including a model for H$_2$ formation can qualitatively change this picture, especially at very high column densities. 

\subsubsection{Variations on {\sc ref} }

\begin{itemize}

\item{\bf SN Feedback and Metal-Line Cooling (Fig.~\ref{fig:sn}):}
The absence of metal-line cooling (model {\sc nozcool}) creates a deficit of strong DLAs while the lack of galactic outflows (model {\sc nosn\_nozcool}) increases $\fN$ by as much as an order of magnitude. When we correct for H$_2$, the same pattern is present up to $\NHI = 10^{21.5} {\rm cm^{-2}}$ but at higher $\NHI$ the dense gas not driven out by winds becomes molecular and turns the surplus in {\sc nosn\_nozcool} into a deficit.

\item{\bf Strong Feedback (Fig.~\ref{fig:agn}):} 
Compared to model {\sc ref}, models with increased feedback efficiency ({\sc agn}, {\sc dblimf}, and {\sc wml4}) all have a deficit in $\fN$ at column densities where the ISM makes a majority contribution (i.e., $\NHI >= 10^{21} {\rm cm^{-2}}$).   The suppression is more effective at higher column densities in models which preferably inject energy into high-mass halos ({\sc agn}, {\sc dblimf}).  Including a correction for H$_2$ changes the amount of suppression but does not lead to a surplus.  

\item{\bf Mass Loading vs. Launch Velocity at Constant Energy per Unit Stellar Mass (Fig.~\ref{fig:wml}):}

For models in which feedback energy per unit stellar mass formed, $\xi$, is kept constant, a lower launch velocity, $v_{\rm w}$, implies a larger mass loading, $\eta$. The abundance of systems below $\NHI \approx 10^{21.25} {\rm cm^{-2}}$ and above $\NHI \approx 10^{22.25} {\rm cm^{-2}}$ is a  monotonically decreasing function of $v_{\rm w}$.  Between these two column densities the ordering of models becomes mixed.  As column density increases through the DLA range, the dominant contribution to $\fN$ first comes from halo gas in low-mass halos, then from the ISM in low-mass halos, then from the ISM in high-mass halos (Table \ref{tab:fN_contributions}). The scaling of $\fN$ for these models is set by the scaling of the halo gas fraction, $f_{\rm hg}$, and the ISM gas fraction, $f_{\rm ISM}$, with halo mass.  The highest mass halos, requiring the largest launch velocities for effective feedback, contribute to the highest column density DLAs and, for these halos, $f_{\rm ISM}$ is larger for smaller $v_{\rm w}$.  At the intermediate DLA column densities, a larger range of halo masses contribute to $\fN$ and models with lower launch velocities can be more effective at reducing the abundance of systems. At the lowest DLA column densities, $f_{\rm hg}$ in low-mass halos becomes more important in determining $\fN$ than $f_{\rm ISM}$.    As long as $v_{\rm w}$ exceeds the threshold for gas to escape the ISM, models with lower $v_{\rm w}$ will increase $f_{\rm hg}$ more efficiently.  In the low-mass halos ($10^9 < M_{\rm halo} / M_{\sun} < 10^{11}$) which determine $\fN$ near the DLA threshold, all values of $v_{\rm w}$ we explored were sufficient.   Including a correction for H$_2$ enhances these trends up to the column densities for which the predicted $\fN$ is truncated. 

\item{\bf Environmentally Dependent Mass Loading and Launch Velocity (Fig.~\ref{fig:whv}):}
We examined two models with increased wind launch velocities $v_{\rm w}$ in high-mass halos. In model {\sc wdens}, $v_{\rm w}$ was proportional to the speed of sound in the ISM and the wind mass loading scaled such that the energy input per unit stellar mass formed was fixed.  In {\sc wvcirc}, $v_{\rm w}$ scaled with halo circular velocity and the wind mass loading scaled such that the momentum input per unit stellar mass formed was fixed.  Both produced patterns similar to model {\sc wml1v848} but had more suppression of the highest column density systems.   This indicates that high velocity winds, even at very low {\em initial} mass loading can entrain ISM material and effectively suppress high column density systems.  Including corrections for H$_2$ in these models results in a dramatic decrease in the number of DLAs above $\NHI \sim 10^{21.5} {\rm cm^{-2}}$ indicating that these high velocity winds increase the halo mass corresponding to a fixed \ion{H}{I} column density, and hence the pressure in the ISM, which then turns molecular.

\item {\bf ISM Equation of State (Fig.~\ref{fig:eos}):}  The effective EOS for star forming gas only affects the abundance of DLAs appreciably when a correction for H$_2$ is included.  For those models, higher pressures in the ISM lead to fewer DLAs and vice versa.

\item{\bf Cosmological Parameters (Fig.~\ref{fig:mill}):}
The greater density of matter and baryons as well as the larger value of $\sigma_8$  in {\sc mill} produce a constant positive offset in $\fN$ relative to {\sc wml4} for $\NHI < 10^{21.25} {\rm cm^{-2}}$.  This effect is independent of the inclusion of H$_2$.  Because {\sc mill} and {\sc wml4} have the same feedback prescription, they have very similar relationships between halo mass and \ion{H}{I} cross-section.  This indicates that differences between the two models are driven by the halo mass function and that it may be possible to use the abundance of DLAs to measure cosmological parameters and the growth of structure.

\item{\bf Star Formation Law (Fig.~\ref{fig:sf}):}
Models with increased star formation efficiency ({\sc sfamplx3}, {\sc sfslope1p75}) have decreased ISM gas fractions due to self-regulation.  This causes a deficit of strong DLAs.  When a correction for H$_2$ is included, the effect in models with low ISM gas fractions is not as strong as it is in model {\sc ref}.  This causes the deficit of DLAs to become a surplus at the highest column densities.  The lower abundance of H$_2$ also truncates $\fN$ at higher column densities than in {\sc ref}.

\item{\bf Timing of Reionization (Fig.~\ref{fig:reion}):}
Models that contain a UV background are very close to {\sc ref} regardless of when that background was turned on (we only considered $z_{\rm reion} \ge 6$).  Model {\sc noreion} with no UV background at all has a constant offset from {\sc ref} until ISM (i.e., totally self-shielded) gas dominates $\fN$ at which point the results agree with {\sc ref}.  These trends are independent of the inclusion of H$_2$.


\end{itemize}

In summary, we find that the \ion{H}{I} CDDF is robust to changes in sub-grid physics models in the LLS column density range and relatively sensitive to them for column densities where the ISM makes a majority contribution to $\fN$ (i.e., the column density range of strong DLAs).  \cite{Rahmati_13_str} showed that the normalization of the CDDF in the LLS column density range changes by less than 0.2 dex with the inclusion of local sources for $z \le 3$ although they find larger changes at higher column densities and redshifts.   The dependencies of $\fN$ on sub-grid physics in the DLA range can, to a large extent, be understood in terms of self-regulated star formation in which the ISM gas fraction adjusts itself until the outflow rate from feedback balances the accretion rate. This suggests that the statistics of relatively strong DLAs can be a valuable resource to constrain sub-grid models. We also showed that DLA statistics are sensitive to the values of some cosmological parameters in a range of column densities not strongly affected by feedback and well constrained observationally.


\section*{Acknowledgments}
We are grateful to all members of the \owls collaboration for their contributions.  The \owls simulations were run on Stella, the {\sc lofar} BlueGene/L system in Groningen, the Cosmology Machine at the ICC which is part of the DiRAC Facility jointly funded by STFC, the Large Facilities Capital Fund of BIS, and Durham University, as part of the Virgo Consortium research programme, and on Darwin in Cambridge. 

This work was sponsored by the National Computing Facilities Foundation (NCF) for the use of supercomputer facilities, with financial support from the Netherlands Organization for Scientific Research (NWO), also through a VIDI grant. The research leading to these results has received funding from the European Research Council under the European Union’s Seventh Framework Programme (FP7/2007-2013) / ERC Grant agreement 278594-GasAroundGalaxies and from the Marie Curie Training Network CosmoComp (PITN-GA-2009-238356). 


\label{lastpage}

\bibliographystyle{mnras}	
\bibliography{paper}	

\begin{thebibliography}{94}
\expandafter\ifx\csname natexlab\endcsname\relax\def\natexlab#1{#1}\fi

\bibitem[{Aguirre} et~al.(2008){Aguirre}, {Dow-Hygelund}, {Schaye} \&
  {Theuns}]{Aguirre_08}
{Aguirre} A., {Dow-Hygelund} C., {Schaye} J., {Theuns} T., 2008, \apj, 689, 851

\bibitem[{Altay} \& {Theuns}(2013)]{Altay_13}
{Altay} G., {Theuns} T., 2013, \mnras, in press

\bibitem[{Altay} et~al.(2011){Altay}, {Theuns}, {Schaye}, {Crighton} \& {Dalla
  Vecchia}]{Altay_11}
{Altay} G., {Theuns} T., {Schaye} J., {Crighton} N.~H.~M., {Dalla Vecchia} C.,
  2011, \apjl, 737, L37

\bibitem[{Bahcall} \& {Peebles}(1969)]{Bahcall_69}
{Bahcall} J.~N., {Peebles} P.~J.~E., 1969, \apjl, 156, L7

\bibitem[{Battisti} et~al.(2012){Battisti}, {Meiring}, {Tripp}
  et~al.]{Battisti_12}
{Battisti} A.~J., {Meiring} J.~D., {Tripp} T.~M., et~al., 2012, \apj, 744, 93

\bibitem[{Becker} \& {Bolton}(2013)]{Becker_13}
{Becker} G.~D., {Bolton} J.~S., 2013, ArXiv e-prints, astro-ph.CO, 1307.2259

\bibitem[{Becker} et~al.(2007){Becker}, {Rauch} \& {Sargent}]{Becker_07}
{Becker} G.~D., {Rauch} M., {Sargent} W.~L.~W., 2007, \apj, 662, 72

\bibitem[{Benson}(2010)]{Benson_10}
{Benson} A.~J., 2010, \physrep, 495, 33

\bibitem[{Bird} et~al.(2013){Bird}, {Vogelsberger}, {Sijacki}, {Zaldarriaga},
  {Springel} \& {Hernquist}]{Bird_13}
{Bird} S., {Vogelsberger} M., {Sijacki} D., {Zaldarriaga} M., {Springel} V.,
  {Hernquist} L., 2013, \mnras, 429, 3341

\bibitem[{Blitz} \& {Rosolowsky}(2006)]{Blitz_06}
{Blitz} L., {Rosolowsky} E., 2006, \apj, 650, 933

\bibitem[{Bolton} \& {Haehnelt}(2007)]{Bolton_07}
{Bolton} J.~S., {Haehnelt} M.~G., 2007, \mnras, 382, 325

\bibitem[{Booth} \& {Schaye}(2009)]{Booth_09}
{Booth} C.~M., {Schaye} J., 2009, \mnras, 398, 53

\bibitem[{Brook} et~al.(2011){Brook}, {Governato}, {Ro{\v s}kar}
  et~al.]{Brook_11}
{Brook} C.~B., {Governato} F., {Ro{\v s}kar} R., et~al., 2011, \mnras, 415,
  1051

\bibitem[{Bryan} et~al.(1999){Bryan}, {Machacek}, {Anninos} \&
  {Norman}]{Bryan_99}
{Bryan} G.~L., {Machacek} M., {Anninos} P., {Norman} M.~L., 1999, \apj, 517, 13

\bibitem[{Carswell} et~al.(1984){Carswell}, {Morton}, {Smith}, {Stockton},
  {Turnshek} \& {Weymann}]{Carswell_84}
{Carswell} R.~F., {Morton} D.~C., {Smith} M.~G., {Stockton} A.~N., {Turnshek}
  D.~A., {Weymann} R.~J., 1984, \apj, 278, 486

\bibitem[{Cen}(2012)]{Cen_12}
{Cen} R., 2012, \apj, 748, 121

\bibitem[{Chabrier}(2003)]{Chabrier_03}
{Chabrier} G., 2003, \pasp, 115, 763

\bibitem[{Cowie} et~al.(1995){Cowie}, {Songaila}, {Kim} \& {Hu}]{Cowie_95}
{Cowie} L.~L., {Songaila} A., {Kim} T.-S., {Hu} E.~M., 1995, \aj, 109, 1522

\bibitem[{Creasey} et~al.(2013){Creasey}, {Theuns} \& {Bower}]{Creasey_13}
{Creasey} P., {Theuns} T., {Bower} R.~G., 2013, \mnras, 429, 1922

\bibitem[{Dalla Vecchia} \& {Schaye}(2008)]{DallaVecchia_08}
{Dalla Vecchia} C., {Schaye} J., 2008, \mnras, 387, 1431

\bibitem[{Dalla Vecchia} \& {Schaye}(2012)]{DallaVecchia_12}
{Dalla Vecchia} C., {Schaye} J., 2012, \mnras, 426, 140

\bibitem[{Dav{\'e}} et~al.(2011){Dav{\'e}}, {Finlator} \&
  {Oppenheimer}]{Dave_11}
{Dav{\'e}} R., {Finlator} K., {Oppenheimer} B.~D., 2011, \mnras, 416, 1354

\bibitem[{Duffy} et~al.(2012){Duffy}, {Meyer}, {Staveley-Smith}
  et~al.]{Duffy_12}
{Duffy} A.~R., {Meyer} M.~J., {Staveley-Smith} L., et~al., 2012, \mnras, 426,
  3385

\bibitem[{Erkal} et~al.(2012){Erkal}, {Gnedin} \& {Kravtsov}]{Erkal_12}
{Erkal} D., {Gnedin} N.~Y., {Kravtsov} A.~V., 2012, \apj, 761, 54

\bibitem[{Faucher-Gigu{\`e}re} et~al.(2008){Faucher-Gigu{\`e}re}, {Lidz},
  {Hernquist} \& {Zaldarriaga}]{FGCA_08}
{Faucher-Gigu{\`e}re} C.-A., {Lidz} A., {Hernquist} L., {Zaldarriaga} M., 2008,
  \apjl, 682, L9

\bibitem[{Ferland} et~al.(1998){Ferland}, {Korista}, {Verner}, {Ferguson},
  {Kingdon} \& {Verner}]{Ferland_89}
{Ferland} G.~J., {Korista} K.~T., {Verner} D.~A., {Ferguson} J.~W., {Kingdon}
  J.~B., {Verner} E.~M., 1998, \pasp, 110, 761

\bibitem[{Fumagalli} et~al.(2011){Fumagalli}, {Prochaska}, {Kasen}, {Dekel},
  {Ceverino} \& {Primack}]{Fumagalli_11}
{Fumagalli} M., {Prochaska} J.~X., {Kasen} D., {Dekel} A., {Ceverino} D.,
  {Primack} J.~R., 2011, \mnras, 418, 1796

\bibitem[{Gnedin} \& {Hui}(1998)]{Gnedin_98}
{Gnedin} N.~Y., {Hui} L., 1998, \mnras, 296, 44

\bibitem[{Gnedin} \& {Kravtsov}(2010)]{2010ApJ...714..287G}
{Gnedin} N.~Y., {Kravtsov} A.~V., 2010, \apj, 714, 287

\bibitem[{Haardt} \& {Madau}(2001)]{HM01}
{Haardt} F., {Madau} P., 2001, in { Clusters of Galaxies and the High Redshift
  Universe Observed in X-rays\/}, edited by {D.~M.~Neumann \& J.~T.~V.~Tran}

\bibitem[{Haardt} \& {Madau}(2012)]{HM12}
{Haardt} F., {Madau} P., 2012, \apj, 746, 125

\bibitem[{Haas} et~al.(2012{\natexlab{a}}){Haas}, {Schaye}, {Booth}
  et~al.]{Haas_12_feed}
{Haas} M.~R., {Schaye} J., {Booth} C.~M., et~al., 2012{\natexlab{a}}, ArXiv
  e-prints, astro-ph.CO, 1211.1021

\bibitem[{Haas} et~al.(2012{\natexlab{b}}){Haas}, {Schaye}, {Booth}
  et~al.]{Haas_12_other}
{Haas} M.~R., {Schaye} J., {Booth} C.~M., et~al., 2012{\natexlab{b}}, ArXiv
  e-prints, astro-ph.CO, 1211.3120

\bibitem[{Haehnelt} et~al.(1998){Haehnelt}, {Steinmetz} \&
  {Rauch}]{Haehnelt_98}
{Haehnelt} M.~G., {Steinmetz} M., {Rauch} M., 1998, \apj, 495, 647

\bibitem[{Heckman} et~al.(1990){Heckman}, {Armus} \& {Miley}]{Heckman_90}
{Heckman} T.~M., {Armus} L., {Miley} G.~K., 1990, \apjs, 74, 833

\bibitem[{Katz} et~al.(1996){Katz}, {Weinberg}, {Hernquist} \&
  {Miralda-Escude}]{Katz_96}
{Katz} N., {Weinberg} D.~H., {Hernquist} L., {Miralda-Escude} J., 1996, \apjl,
  457, L57

\bibitem[{Kennicutt}(1998)]{Kennicut_98}
{Kennicutt} Jr. R.~C., 1998, \apj, 498, 541

\bibitem[{Kim} et~al.(2002){Kim}, {Carswell}, {Cristiani}, {D'Odorico} \&
  {Giallongo}]{Kim_02}
{Kim} T.-S., {Carswell} R.~F., {Cristiani} S., {D'Odorico} S., {Giallongo} E.,
  2002, \mnras, 335, 555

\bibitem[{Kim} et~al.(2013){Kim}, {Partl}, {Carswell} \& {M{\"u}ller}]{Kim_13}
{Kim} T.-S., {Partl} A.~M., {Carswell} R.~F., {M{\"u}ller} V., 2013, \aap, 552,
  A77

\bibitem[{Kohler} \& {Gnedin}(2007)]{Kohler_07}
{Kohler} K., {Gnedin} N.~Y., 2007, \apj, 655, 685

\bibitem[{Komatsu} et~al.(2011){Komatsu}, {Smith}, {Dunkley}
  et~al.]{Komatsu_11}
{Komatsu} E., {Smith} K.~M., {Dunkley} J., et~al., 2011, \apjs, 192, 18

\bibitem[{Krumholz} et~al.(2009){Krumholz}, {McKee} \&
  {Tumlinson}]{2009ApJ...693..216K}
{Krumholz} M.~R., {McKee} C.~F., {Tumlinson} J., 2009, \apj, 693, 216

\bibitem[{Lanzetta} et~al.(1991){Lanzetta}, {Wolfe}, {Turnshek}, {Lu},
  {McMahon} \& {Hazard}]{Lanzetta_91}
{Lanzetta} K.~M., {Wolfe} A.~M., {Turnshek} D.~A., {Lu} L., {McMahon} R.~G.,
  {Hazard} C., 1991, \apjs, 77, 1

\bibitem[{McQuinn} et~al.(2011){McQuinn}, {Oh} \&
  {Faucher-Gigu{\`e}re}]{McQuinn_11}
{McQuinn} M., {Oh} S.~P., {Faucher-Gigu{\`e}re} C.-A., 2011, \apj, 743, 82

\bibitem[{Meiksin}(2009)]{Meiksin_09}
{Meiksin} A.~A., 2009, Reviews of Modern Physics, 81, 1405

\bibitem[{Miralda-Escud{\'e}}(2005)]{MiraldaEscude_05}
{Miralda-Escud{\'e}} J., 2005, \apjl, 620, L91

\bibitem[{M{\o}ller} et~al.(2013){M{\o}ller}, {Fynbo}, {Ledoux} \&
  {Nilsson}]{Moller_13}
{M{\o}ller} P., {Fynbo} J.~P.~U., {Ledoux} C., {Nilsson} K.~K., 2013, \mnras,
  430, 2680

\bibitem[{Noterdaeme} et~al.(2012){Noterdaeme}, {Petitjean}, {Carithers}
  et~al.]{Noterdaeme_12}
{Noterdaeme} P., {Petitjean} P., {Carithers} W.~C., et~al., 2012, \aap, 547, L1

\bibitem[{Noterdaeme} et~al.(2009){Noterdaeme}, {Petitjean}, {Ledoux} \&
  {Srianand}]{Noterdaeme_09}
{Noterdaeme} P., {Petitjean} P., {Ledoux} C., {Srianand} R., 2009, \aap, 505,
  1087

\bibitem[{Okamoto} et~al.(2008){Okamoto}, {Gao} \& {Theuns}]{Okamoto_08}
{Okamoto} T., {Gao} L., {Theuns} T., 2008, \mnras, 390, 920

\bibitem[{O'Meara} et~al.(2007){O'Meara}, {Prochaska}, {Burles}, {Prochter},
  {Bernstein} \& {Burgess}]{Omeara_07}
{O'Meara} J.~M., {Prochaska} J.~X., {Burles} S., {Prochter} G., {Bernstein}
  R.~A., {Burgess} K.~M., 2007, \apj, 656, 666

\bibitem[{O'Meara} et~al.(2013){O'Meara}, {Prochaska}, {Worseck}, {Chen} \&
  {Madau}]{Omeara_13}
{O'Meara} J.~M., {Prochaska} J.~X., {Worseck} G., {Chen} H.-W., {Madau} P.,
  2013, \apj, 765, 137

\bibitem[{Patra} et~al.(2013){Patra}, {Chengalur} \& {Begum}]{Patra_13}
{Patra} N.~N., {Chengalur} J.~N., {Begum} A., 2013, \mnras, 429, 1596

\bibitem[{Pawlik} et~al.(2009){Pawlik}, {Schaye} \& {van
  Scherpenzeel}]{Pawlik_09}
{Pawlik} A.~H., {Schaye} J., {van Scherpenzeel} E., 2009, \mnras, 394, 1812

\bibitem[{P{\'e}roux} et~al.(2001){P{\'e}roux}, {Storrie-Lombardi}, {McMahon},
  {Irwin} \& {Hook}]{Peroux_01}
{P{\'e}roux} C., {Storrie-Lombardi} L.~J., {McMahon} R.~G., {Irwin} M., {Hook}
  I.~M., 2001, \aj, 121, 1799

\bibitem[{Petitjean} et~al.(1993){Petitjean}, {Webb}, {Rauch}, {Carswell} \&
  {Lanzetta}]{Petitjean_93}
{Petitjean} P., {Webb} J.~K., {Rauch} M., {Carswell} R.~F., {Lanzetta} K.,
  1993, \mnras, 262, 499

\bibitem[{Pettini} et~al.(2001){Pettini}, {Shapley}, {Steidel}
  et~al.]{Pettini_01}
{Pettini} M., {Shapley} A.~E., {Steidel} C.~C., et~al., 2001, \apj, 554, 981

\bibitem[{Planck Collaboration} et~al.(2013){Planck Collaboration}, {Ade},
  {Aghanim} et~al.]{Planck_13}
{Planck Collaboration}, {Ade} P.~A.~R., {Aghanim} N., et~al., 2013, ArXiv
  e-prints, astro-ph.CO, 1303.5076

\bibitem[{Pontzen} et~al.(2008){Pontzen}, {Governato}, {Pettini}
  et~al.]{Pontzen_08}
{Pontzen} A., {Governato} F., {Pettini} M., et~al., 2008, \mnras, 390, 1349

\bibitem[{Prochaska} et~al.(2005){Prochaska}, {Herbert-Fort} \&
  {Wolfe}]{Prochaska_05}
{Prochaska} J.~X., {Herbert-Fort} S., {Wolfe} A.~M., 2005, \apj, 635, 123

\bibitem[{Prochaska} et~al.(2010){Prochaska}, {O'Meara} \&
  {Worseck}]{Prochaska_10}
{Prochaska} J.~X., {O'Meara} J.~M., {Worseck} G., 2010, \apj, 718, 392

\bibitem[{Prochaska} \& {Wolfe}(2009)]{Prochaska_09}
{Prochaska} J.~X., {Wolfe} A.~M., 2009, \apj, 696, 1543

\bibitem[{Rahmati} et~al.(2013{\natexlab{a}}){Rahmati}, {Pawlik}, {Raicevic} \&
  {Schaye}]{Rahmati_13_evo}
{Rahmati} A., {Pawlik} A.~H., {Raicevic} M., {Schaye} J., 2013{\natexlab{a}},
  \mnras, 430, 2427

\bibitem[{Rahmati} et~al.(2013{\natexlab{b}}){Rahmati}, {Schaye}, {Pawlik} \&
  {Raicevic}]{Rahmati_13_str}
{Rahmati} A., {Schaye} J., {Pawlik} A.~H., {Raicevic} M., 2013{\natexlab{b}},
  \mnras, 431, 2261

\bibitem[{Rakic} et~al.(2012){Rakic}, {Schaye}, {Steidel} \& {Rudie}]{Rakic_12}
{Rakic} O., {Schaye} J., {Steidel} C.~C., {Rudie} G.~C., 2012, \apj, 751, 94

\bibitem[{Rauch}(1998)]{Rauch_98}
{Rauch} M., 1998, \araa, 36, 267

\bibitem[{Ribaudo} et~al.(2011{\natexlab{a}}){Ribaudo}, {Lehner} \&
  {Howk}]{Ribaudo_11_Census}
{Ribaudo} J., {Lehner} N., {Howk} J.~C., 2011{\natexlab{a}}, \apj, 736, 42

\bibitem[{Ribaudo} et~al.(2011{\natexlab{b}}){Ribaudo}, {Lehner}, {Howk}
  et~al.]{Ribaudo_11_ColdAcc}
{Ribaudo} J., {Lehner} N., {Howk} J.~C., et~al., 2011{\natexlab{b}}, \apj, 743,
  207

\bibitem[{Rudie} et~al.(2013){Rudie}, {Steidel}, {Shapley} \&
  {Pettini}]{Rudie_13}
{Rudie} G.~C., {Steidel} C.~C., {Shapley} A.~E., {Pettini} M., 2013, \apj, 769,
  146

\bibitem[{Scannapieco} et~al.(2012){Scannapieco}, {Wadepuhl}, {Parry}
  et~al.]{Scannapieco_12}
{Scannapieco} C., {Wadepuhl} M., {Parry} O.~H., et~al., 2012, \mnras, 423, 1726

\bibitem[{Schaye}(2001)]{Schaye_01}
{Schaye} J., 2001, \apjl, 562, L95

\bibitem[{Schaye}(2004)]{Schaye_04}
{Schaye} J., 2004, \apj, 609, 667

\bibitem[{Schaye}(2006)]{Schaye_06}
{Schaye} J., 2006, \apj, 643, 59

\bibitem[{Schaye} et~al.(2003){Schaye}, {Aguirre}, {Kim}, {Theuns}, {Rauch} \&
  {Sargent}]{Schaye_03}
{Schaye} J., {Aguirre} A., {Kim} T.-S., {Theuns} T., {Rauch} M., {Sargent}
  W.~L.~W., 2003, \apj, 596, 768

\bibitem[{Schaye} \& {Dalla Vecchia}(2008)]{Schaye_08}
{Schaye} J., {Dalla Vecchia} C., 2008, \mnras, 383, 1210

\bibitem[{Schaye} et~al.(2010){Schaye}, {Dalla Vecchia}, {Booth}
  et~al.]{Schaye_10}
{Schaye} J., {Dalla Vecchia} C., {Booth} C.~M., et~al., 2010, \mnras, 402, 1536

\bibitem[{Schaye} et~al.(2000){Schaye}, {Theuns}, {Rauch}, {Efstathiou} \&
  {Sargent}]{Schaye_00}
{Schaye} J., {Theuns} T., {Rauch} M., {Efstathiou} G., {Sargent} W.~L.~W.,
  2000, \mnras, 318, 817

\bibitem[{Shen} et~al.(2013){Shen}, {Madau}, {Guedes}, {Mayer}, {Prochaska} \&
  {Wadsley}]{Shen_13}
{Shen} S., {Madau} P., {Guedes} J., {Mayer} L., {Prochaska} J.~X., {Wadsley}
  J., 2013, \apj, 765, 89

\bibitem[{Spergel} et~al.(2007){Spergel}, {Bean}, {Dor{\'e}}
  et~al.]{2007ApJS..170..377S}
{Spergel} D.~N., {Bean} R., {Dor{\'e}} O., et~al., 2007, \apjs, 170, 377

\bibitem[{Springel}(2005)]{Springel_05}
{Springel} V., 2005, \mnras, 364, 1105

\bibitem[{Springel}(2010)]{Springel_10}
{Springel} V., 2010, \mnras, 401, 791

\bibitem[{Steidel} et~al.(2010){Steidel}, {Erb}, {Shapley} et~al.]{Steidel_10}
{Steidel} C.~C., {Erb} D.~K., {Shapley} A.~E., et~al., 2010, \apj, 717, 289

\bibitem[{Storrie-Lombardi} \& {Wolfe}(2000)]{Storrie-Lombardi_00}
{Storrie-Lombardi} L.~J., {Wolfe} A.~M., 2000, \apj, 543, 552

\bibitem[{Theuns} et~al.(2000){Theuns}, {Schaye} \& {Haehnelt}]{Theuns_00}
{Theuns} T., {Schaye} J., {Haehnelt} M.~G., 2000, \mnras, 315, 600

\bibitem[{Theuns} et~al.(2002){Theuns}, {Viel}, {Kay}, {Schaye}, {Carswell} \&
  {Tzanavaris}]{Theuns_02}
{Theuns} T., {Viel} M., {Kay} S., {Schaye} J., {Carswell} R.~F., {Tzanavaris}
  P., 2002, \apjl, 578, L5

\bibitem[{Tytler}(1987)]{Tytler_87}
{Tytler} D., 1987, \apj, 321, 49

\bibitem[{van de Voort} \& {Schaye}(2012)]{VanDeVoort_Schaye_12}
{van de Voort} F., {Schaye} J., 2012, \mnras, 423, 2991

\bibitem[{van de Voort} et~al.(2012){van de Voort}, {Schaye}, {Altay} \&
  {Theuns}]{VanDeVoort_12}
{van de Voort} F., {Schaye} J., {Altay} G., {Theuns} T., 2012, \mnras, 421,
  2809

\bibitem[{Viel} et~al.(2013){Viel}, {Schaye} \& {Booth}]{Viel_13}
{Viel} M., {Schaye} J., {Booth} C.~M., 2013, \mnras, 429, 1734

\bibitem[{Wiersma} et~al.(2009{\natexlab{a}}){Wiersma}, {Schaye} \&
  {Smith}]{Wiersma_09_cool}
{Wiersma} R.~P.~C., {Schaye} J., {Smith} B.~D., 2009{\natexlab{a}}, \mnras,
  393, 99

\bibitem[{Wiersma} et~al.(2009{\natexlab{b}}){Wiersma}, {Schaye}, {Theuns},
  {Dalla Vecchia} \& {Tornatore}]{Wiersma_09_chem}
{Wiersma} R.~P.~C., {Schaye} J., {Theuns} T., {Dalla Vecchia} C., {Tornatore}
  L., 2009{\natexlab{b}}, \mnras, 399, 574

\bibitem[{Wolfe} et~al.(2005){Wolfe}, {Gawiser} \& {Prochaska}]{Wolfe_05}
{Wolfe} A.~M., {Gawiser} E., {Prochaska} J.~X., 2005, \araa, 43, 861

\bibitem[{Yajima} et~al.(2012){Yajima}, {Choi} \& {Nagamine}]{Yajima_12}
{Yajima} H., {Choi} J.-H., {Nagamine} K., 2012, \mnras, 427, 2889

\bibitem[{Zafar} et~al.(2013){Zafar}, {Peroux}, {Popping}, {Milliard},
  {Deharveng} \& {Frank}]{Zafar_13}
{Zafar} T., {Peroux} C., {Popping} A., {Milliard} B., {Deharveng} J.-M.,
  {Frank} S., 2013, ArXiv e-prints, astro-ph.CO, 1307.0602

\end{thebibliography}


\appendix

\section{Hydrodynamics Solvers}

\begin{figure}
  \begin{center}
    \includegraphics[width=0.49\textwidth]{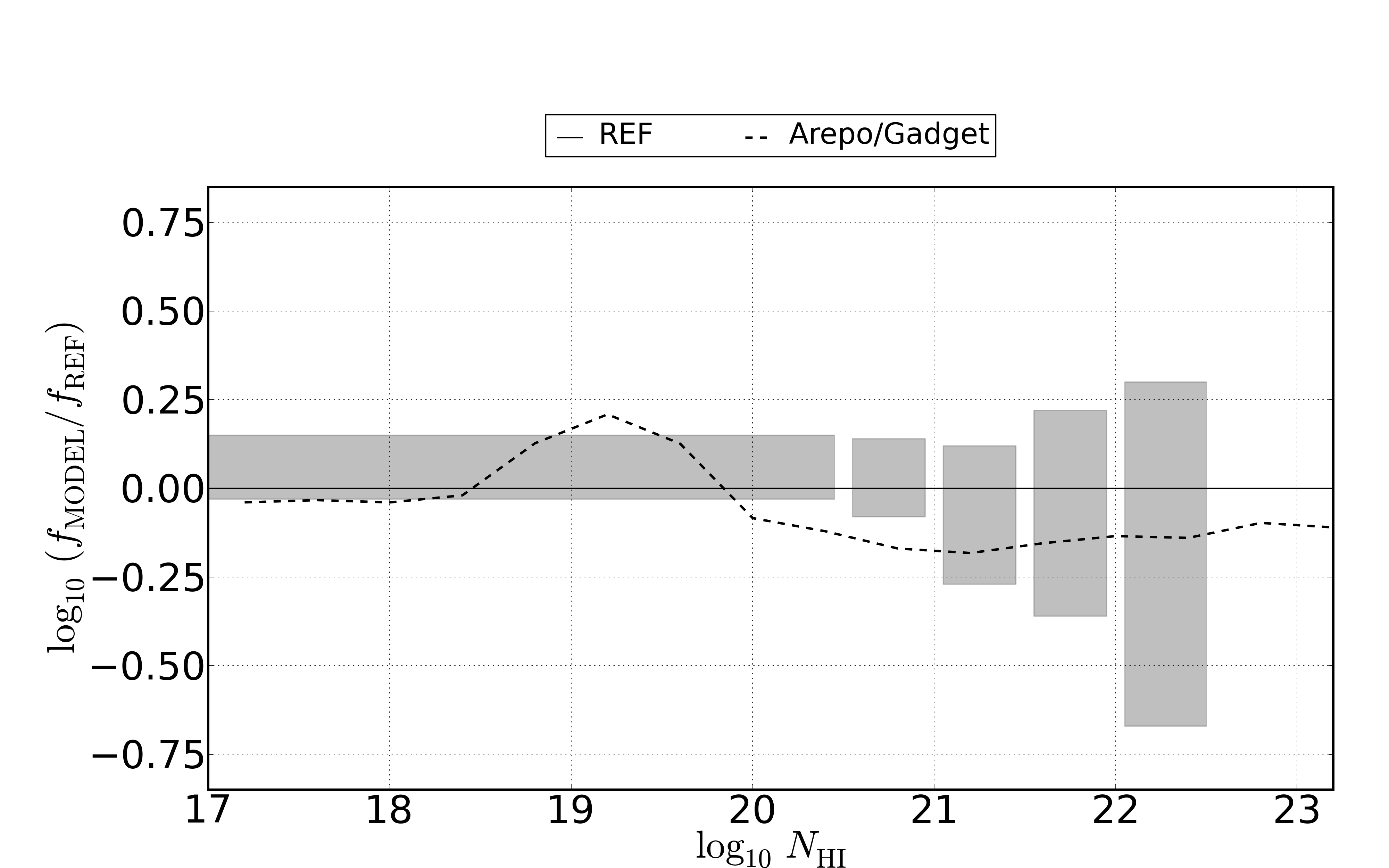}
  \end{center}
  \caption{A reproduction of Fig. \ref{fig:main_cddf_big} which focuses on data from \protect \cite{Bird_13}.  The {\em dashed black} line is the ratio of $\fN$ in the \arepo and \gadget runs from that work (solid black line in their Fig. 7).  \protect \cite{Bird_13} conclude that the surplus around $\NHI=10^{19} {\rm cm^{-2}}$ is due to their simplified self-shielding correction and the deficit above $\NHI \approx 10^{20} {\rm cm^{-2}}$ is due to lower peak halo densities in \arepo compared to \gadget.  The magnitude of the change due to halo densities is comparable to or smaller than that due to sub-grid variations.}
\label{fig:main_cddf_big_Arepo}
\end{figure}

In this appendix we briefly compare differences in $\fN$ caused by \owls model variations to those due to the choice of hydrodynamics solver.  Fig. \ref{fig:main_cddf_big_Arepo} is in the style of Figure~\ref{fig:main_cddf_big}.  The dashed line labeled \arepo / \gadget is taken from \cite{Bird_13} (solid black line in Figure 7 of their work), who compared results from a simulation performed with {\sc arepo} \citep{Springel_10} to those from a simulation performed using a version of {\sc gadget} \citep{Springel_05}. The two simulations in \cite{Bird_13} are identical in terms of cosmological parameters and sub-grid physics but use different hydrodynamics solvers (moving mesh for {\sc arepo} versus SPH for {\sc gadget}).  To simplify the comparison between the two codes feedback parameters were chosen such that strong outflows were not driven.  Keeping in mind that the sub-grid models used in \cite{Bird_13} are different from any of the \owls\ models, we compare the variation among \owls models (grey boxes) to the ratio from \cite{Bird_13}.

Setting aside the surplus around $\NHI=10^{19} {\rm cm^{-2}}$, which \cite{Bird_13} conclude is due to their simplified self-shielding correction, the magnitude of differences due to the choice of hydrodynamics solver are comparable to or smaller than the differences between \owls variations over most of the $\NHI$ range. 
However, it is interesting to note the deficit above $\NHI \approx 10^{20} {\rm cm^{-2}}$ which is due to lower peak halo densities in \arepo compared to {\sc gadget}.
This change is consistent with the work of \cite{VanDeVoort_12} who found that $\NHI \approx 10^{20} {\rm cm^{-2}}$ is when the ISM begins to make a contribution to $\fN$.  Thus the peak density in gaseous halos as well as the impact of local sources both represent sources of uncertainty if the amplitude of $\fN$ is to be used to constrain cosmological parameters such as $\sigma_8$ (see discussion in \S 6.2.2).  The question of how the introduction of more complicated sub-grid models will affect the results of moving mesh or grid codes remains open.  However, it has been shown, in the context of galaxy formation, that variations in sub-grid physics typically lead to changes at least as large as those due to the choice of hydrodynamics solver \citep{Scannapieco_12}.

\end{document}